\documentclass[prx,aps,showpacs,twocolumn, preprintnumbers,
amsmath,amssymb,superscriptaddress,nofootinbib,longbibliography,nofootinbibjustification=justified,singlelinecheck=false]{revtex4-2}

\usepackage{amsfonts}
\usepackage{graphicx}
\usepackage{hyperref} 
\usepackage{cleveref}
\usepackage{enumitem}
\DeclareMathOperator{\Arg}{Arg}

\usepackage{listings}
\usepackage{color}

\hypersetup{
    hidelinks,
    colorlinks=true,
    breaklinks=true,
    citecolor=SteelBlue,
    filecolor=LimeGreen,
    linkcolor=MediumBlue,
    urlcolor=MediumPurple
}
\usepackage[svgnames]{xcolor}

\newcommand{\ket}[1]{|#1\rangle}

\newcommand{\bra}[1]{ \langle #1 \,|}

\usepackage{lipsum} % for random filler text

\begin{document}

\title{Analytical Fock Representation of Two-Mode Squeezing for Quantum Interference}

\author{Xuemei Gu}
\affiliation{Institut für Festkörpertheorie und Optik, Friedrich-Schiller-Universität Jena, Jena, Germany}
\email{xuemei.gu@uni-jena.de}

\author{Carlos Ruiz-Gonzalez}
\affiliation{Max Planck Institute for the Science of Light, Erlangen, Germany}
%\email{cruizgo@proton.me}

\author{Mario Krenn}
\affiliation{Max Planck Institute for the Science of Light, Erlangen, Germany}
\affiliation{Machine Learning in Science Cluster, Department for Computer Science, Faculty of Science, University of Tuebingen, Tuebingen, Germany}
\email{mario.krenn@uni-tuebingen.de}

\begin{abstract}
%Two-mode squeezing is central to entangled-photon generation and nonlinear interferometry, yet standard perturbative low-gain and Gaussian treatments obscure how photon-number amplitudes interfere, especially in multi-crystal geometries and at high gain. Here, we derive the exact analytic action of the two-mode squeezing operator on arbitrary Fock states to analyze nonlinear interferometers directly in the number basis at arbitrary squeezing strength. Within this framework, we find new physical interpretations of previously known quantum interference effects, and theoretically discover a new and unusual multi-photon interference effect in an experimental four-crystal geometry that could readily be observed in laboratories. Our work provides a compact analytic toolkit and concrete design rules for engineering multi-photon interference, with applications in quantum sensing, precision metrology, and advanced quantum state generation.
Two-mode squeezing is central to entangled-photon generation and nonlinear interferometry, yet standard perturbative low-gain treatments and Gaussian formalisms can obscure the interference of photon-number amplitudes, especially in nonlinear interferometers and at high gain. Here we derive a closed-form Fock-basis expression for the action of the two-mode squeezing operator on arbitrary number states, enabling the direct analysis of nonlinear interferometers in the photon-number basis at arbitrary squeezing strength. Within this framework, we provide intuitive physical interpretations of several known quantum-interference effects and identify a new multi-photon interference phenomenon in a four-crystal geometry that could readily be observed in laboratories. Our work provides a compact analytic toolkit and explicit design rules for engineering multi-photon interference, with applications in quantum sensing, precision metrology, and quantum state generation.
\end{abstract}
\maketitle

\section*{Introduction}
Squeezed light is an important resource in quantum optics. In particular, the two-mode squeezing operator underpins entangled-photon generation by spontaneous parametric down-conversion, and has a broad range of applications in sensing and imaging \cite{yurke19862, chekhova2016nonlinear, ou2020quantum, braunstein2005quantum,zheng2025nonlinear}. Most analyses adopt either (i) a \emph{low-gain}, perturbative treatment that truncates multi-pair creation, or (ii) the \emph{continuous-variable} (Gaussian) formalism that emphasizes quadratures, Bogoliubov transformations, and homodyne detection \cite{braunstein2005quantum}. Both perspectives are powerful and have enabled major advances in quantum metrology and quantum imaging \cite{machado2020optical,kviatkovsky2020microscopy,barreto2022quantum,hochrainer2022quantum,bowen2023quantum,spagnolo2023non}. 

While the perturbative low-gain model drops higher-order emissions by construction, the continuous-variable (Gaussian) formalism, though powerful for quadrature analyses, embeds fixed photon-number amplitudes inside high-order moments. Consequently, the phase-dependent cancellations between concrete emission events -- crucial in modern nonlinear interferometers -- are not directly visible without reverting to a number-state expansion.

The exact action of the two-mode squeezing operator on number states is known in the mathematical-physics literature~\cite{chizhov1993photon, barnett2002methods, kok2010introduction}, but it is not often used as a practical analytic tool for interpreting modern quantum interferometers, where interference among higher-order emission events leads to surprising and often counter-intuitive effects. Prominent examples include frustrated pair creation at low gain~\cite{herzog1994frustrated, jiang2025subjective}, two-boson interference in time and its high-gain cancellation of a seeded single photon pair~\cite{cerf2020two, chen2025two} and its generalizations~\cite{jabbour2021multiparticle}, as well as multi-photon nonlocal interference controlled by undetected photons~\cite{qian2023multiphoton, feng2023chip, wang2025violation}. In such settings, a transparent, exact \emph{Fock-basis} description reveals where the interference comes from and how it changes with gain.

In this work, we derive an exact Fock-basis expression for the action of the two-mode squeezing operator on arbitrary Fock states. We use it for deriving new physical interpretations of numerous photonic nonlinear phenomena, and derive new counterintuitive multi-photon interference that can readily be observed in laboratories. Concretely, (1) we provide a simple physical interpretation of the two-photon interference effect predicted by Cerf and Jabbour ~\cite{cerf2020two}, (2) we reinterpret high-gain phase sensitivity in SU(1,1) interferometers using the Fock basis, (3) we derive the conditions for perfect two-photon destructive interference in three-crystal nonlinear interferometers \cite{jiang2025subjective}, and (4) we find new four-photon interference effects in which perfect cancellation requires asymmetric squeezing conditions, including an unexpected cancellation at zero relative phase analogous to the Cerf-Jabbour effect. Together, these results demonstrate how powerful the analytical Fock-basis representation of the squeezing operator is for reinterpreting known and discovering new complex interference effects in quantum optics.

%Concretely, (1) we find a simple physical interpretation for a recently discovered two-photon interference effect at high-gain by Cerf and Jabbour~\cite{cerf2020two}, (2) we reinterpret a high-gain effect in SU(1,1) interferometers using the Fock basis, (3) we find perfect interference conditions of a recent four-photon non-linear interferometer in the high-gain regime and (4) find a new, counter-intuitive four-photon interference effect that has analogies to the Cerf-Jabbour effect. 

\section*{Analytic Expression and Quantum Interference} 
We begin by deriving analytic expressions for the action of the two-mode squeezing operator in the Fock basis and then apply them to analyze quantum interference across single- and multi-crystal configurations.

\subsection*{Analytic Fock-basis representation of squeezed states}
Consider the two-mode squeezing operator $S_2(\zeta)$
\begin{equation}
S_2(\zeta) = \exp \left(\zeta^* a b  - \zeta a^\dagger b^\dagger\right),\label{eq:s2operator}
\end{equation}
where $\zeta = r e^{i\theta}$ is the complex squeezing parameter, and $a$ ($a^\dagger$) and $b$ ($b^\dagger$) denote the annihilation (creation) operators of the two modes $a$ and $b$. Throughout this work, we write \(S_{ij}(\zeta)\) to indicate that the two-mode squeezing operator acts on modes \(i\) and \(j\); for example, \(S_{ab}(\zeta)\) acts on modes \(a\) and \(b\), while \(S_{cd}(\zeta)\) acts on modes \(c\) and \(d\). Using the disentanglement formula for the $\mathrm{SU}(1,1)$ Lie algebra~\cite{truax1985baker, dasgupta1996disentanglement}, the operator $S_2(\zeta)$ in Eq.~\eqref{eq:s2operator} can be rewritten as
\begin{align}
S_2(\zeta) &=\exp \left(-e^{i\theta} \tanh r \, a^\dagger b^\dagger \right)\nonumber\\
&\quad\times \exp \left( -\ln(\cosh r) \, (a^\dagger a + b^\dagger b + 1) \right)\nonumber\\
&\quad\times \exp \left(e^{-i\theta} \tanh r \, ab \right),\label{eq:S2normal_order}
\end{align}
This normal-ordered form provides a convenient starting point for deriving the action of the squeezing operator on general Fock states. The detailed derivation from Eq.~\eqref{eq:s2operator} to Eq.~\eqref{eq:S2normal_order} is presented in the Appendix.

We apply $S_2(\zeta)$ in Eq.~\eqref{eq:S2normal_order} to a general two-mode Fock state $\ket{p,q}$ (with \(p\) photons in mode \(a\) and \(q\) photons in mode \(b\)), which gives
\begin{align}
\ket{\psi}&=S_2(\zeta)\ket{p, q}\nonumber\\
&=\sum_{k=0}^{\infty} \sum_{n=0}^{\min{(p,q)}}\Biggl[ \frac{(e^{-i\theta}\tanh r)^{n}(-e^{i\theta}\tanh r)^{k}}{(\cosh r)^{p+q-2n+1}}\nonumber\\
&\quad\times \sqrt{\binom{p}{n}\binom{q}{n}\binom{p-n+k}{k}\binom{q-n+k}{k}}\nonumber\\
&\quad \times \,\,\ket{p-n+k, q-n+k}\Biggl]
\label{eq:S2Fockpq}
\end{align}
Although this formula is not frequently discussed explicitly, it appears in the literature~\cite{chizhov1993photon,hu2008two} and can be expressed in a recursive way~\cite{miatto2020fast, jabbour2021multiparticle}. Importantly, we show that it provides a useful framework to describe nonlinear interferometers directly in the Fock basis, enabling both a reinterpretation of known results and the prediction of new interference phenomena.

For the special Fock states $\ket{p,p}$, Eq.~\eqref{eq:S2Fockpq} simplifies to
\begin{align}
\ket{\psi}&=S_2(\zeta)\ket{p, p}\nonumber\\
&=\sum_{k=0}^{\infty} \sum_{n=0}^{p}\Biggl[\frac{(e^{-i\theta}\tanh r)^{n}(-e^{i\theta}\tanh r)^{k}}{(\cosh r)^{2p-2n+1}}\nonumber\\
&\quad\times\binom{p}{n}\binom{p-n+k}{k}\ket{p-n+k, p-n+k}\Biggl]
\label{eq:S2Fockp}
\end{align}
while for the vacuum input $\ket{0,0}$, one can obtain the standard two-mode squeezed vacuum state, which is
\begin{align}
\ket{\psi}&=S_2(\zeta)\ket{0,0}=\sum\limits_{m=0}^{\infty} \frac{(-e^{i \theta}\tanh{r})^{m}}{\cosh{r}}\,\, \ket{m, m} \label{eq:S2Fockvacuum}
\end{align}

In the following, we systematically analyze experimental configurations from a single crystal to four-crystal arrays, showing how Eq.~\eqref{eq:S2Fockpq} describes quantum interference. For each setup, we consider all the squeezing phases are zero (i.e., $\theta=0$) and extract the amplitude for specific output states and identify conditions for interference.

\subsection*{Two-photon interference in a single crystal} 
\begin{figure}[!t]
    \centering
    \includegraphics[width=1\linewidth]{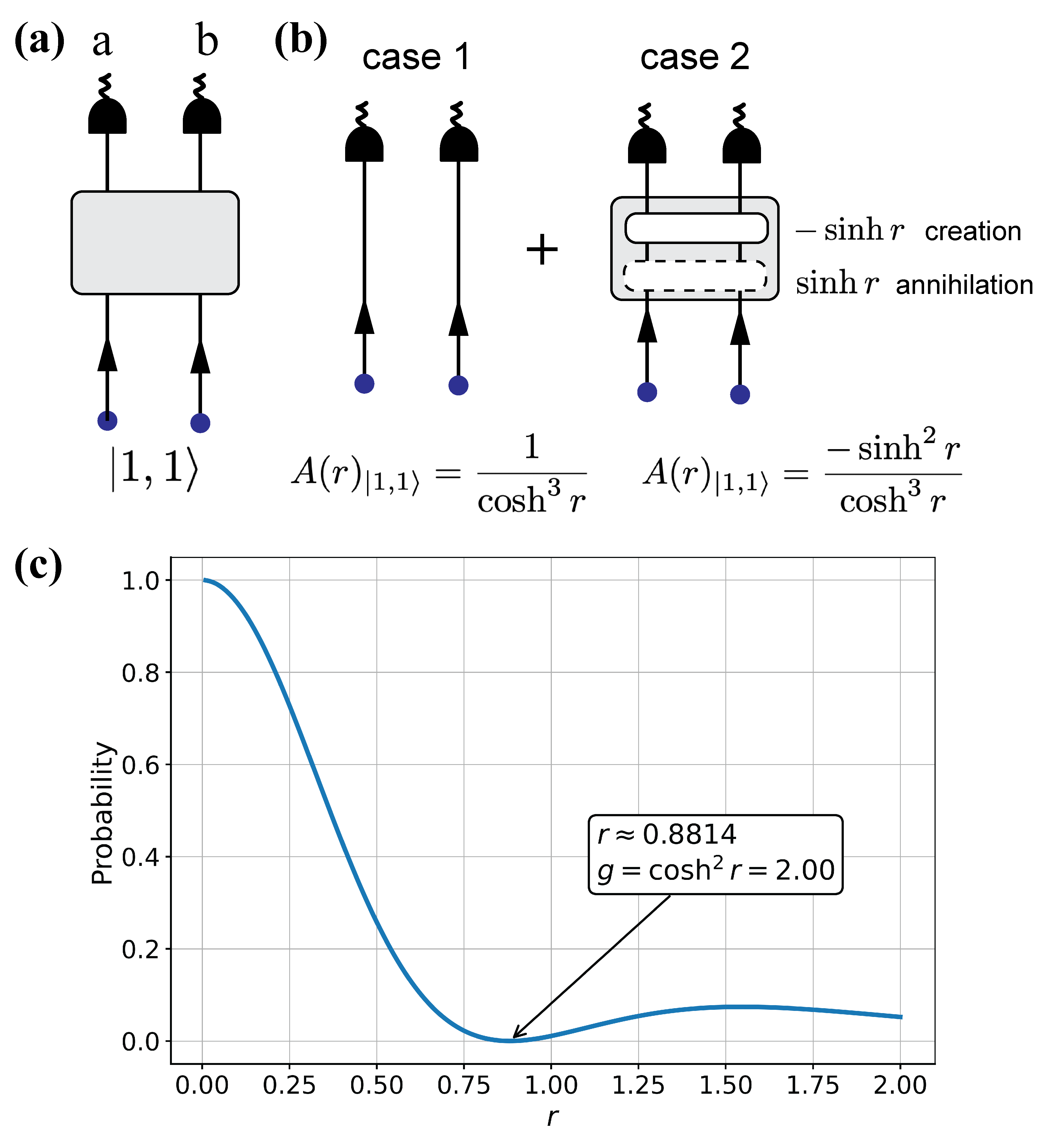}
    \caption{A single nonlinear crystal seeded with two single photons and the two-photon coincidence probability. \textbf{(a)} Two indistinguishable single photons (blue dots) enter a nonlinear crystal and the output detection paths are $a$ and $b$. \textbf{(b)} The photons evolve into a superposition of being either directly transmitted or being reborn in the nonlinear crystal (annihilation followed by creation of an indistinguishable pair). \textbf{(c)} The coincidence probability $P_{\ket{1,1}}$ is plotted versus the squeezing parameter $r$. At $r=\operatorname{arcsinh}(1)\approx 0.88$, the $\ket{1,1}$ amplitude vanishes, in agreement with Ref.~\cite{cerf2020two}.}
    \label{1crystal}
\end{figure}
An interesting form of destructive quantum interference occurs when a single nonlinear crystal is seeded with the input Fock state $\ket{1,1}$ (Fig.~\ref{1crystal} (a)). This effect was theoretically predicted by Cerf and Jabbour as the time-reversed analogue of Hong-Ou-Mandel interference~\cite{cerf2020two} and was recently observed experimentally~\cite{chen2025two}. 

Applying Eq.~\eqref{eq:S2Fockpq} with the input state $\ket{1,1}$ (where $p=q=1$), the amplitude for observing $\ket{1,1}$ at the output is
\begin{align}
A(r)_{\ket{1,1}}=\bra{1,1}S_{2}(r)\ket{1,1}=\frac{1-\sinh^2 r}{\cosh^{3} r}.
\label{eq:singleCrystalInterference}
\end{align}
This amplitude becomes zero at $r=\operatorname{arcsinh}(1)\approx0.88$, e.g., the coincidence rate for detecting $\ket{1,1}$ photons vanishes completely (as shown in Fig.~\ref{1crystal} (c)). With the parametrization $g=\cosh^{2}r$, the same expression can be written as $A_{\ket{1,1}}=(2 - g)/g^{\frac{3}{2}}$, showing perfect destructive interference at $g=2$, exactly matching the prediction of Ref.~\cite{cerf2020two}.
 
The new interpretation of the interference effect uses Eq.~\eqref{eq:singleCrystalInterference}. Physically, the cancellation arises from the interference between two indistinguishable quantum processes (Fig.~\ref{1crystal} (b)). The first process is direct transmission of the input photon pair through the crystal. The second involves annihilation of both input photons followed by creation of a new photon pair within the crystal. Each annihilation and creation contributes an amplitude factor of $\sinh r$ (but with an opposite sign), and their combination leads to a total amplitude of $-\sinh^2 r$ for the second process. Perfect destructive interference occurs when the two components have equal magnitude and opposite phase, which happens when $\sinh^2 r=1$.

\subsection*{Two-photon interference between two crystals}
\begin{figure}[!t]
    \centering
    \includegraphics[width=1\linewidth]{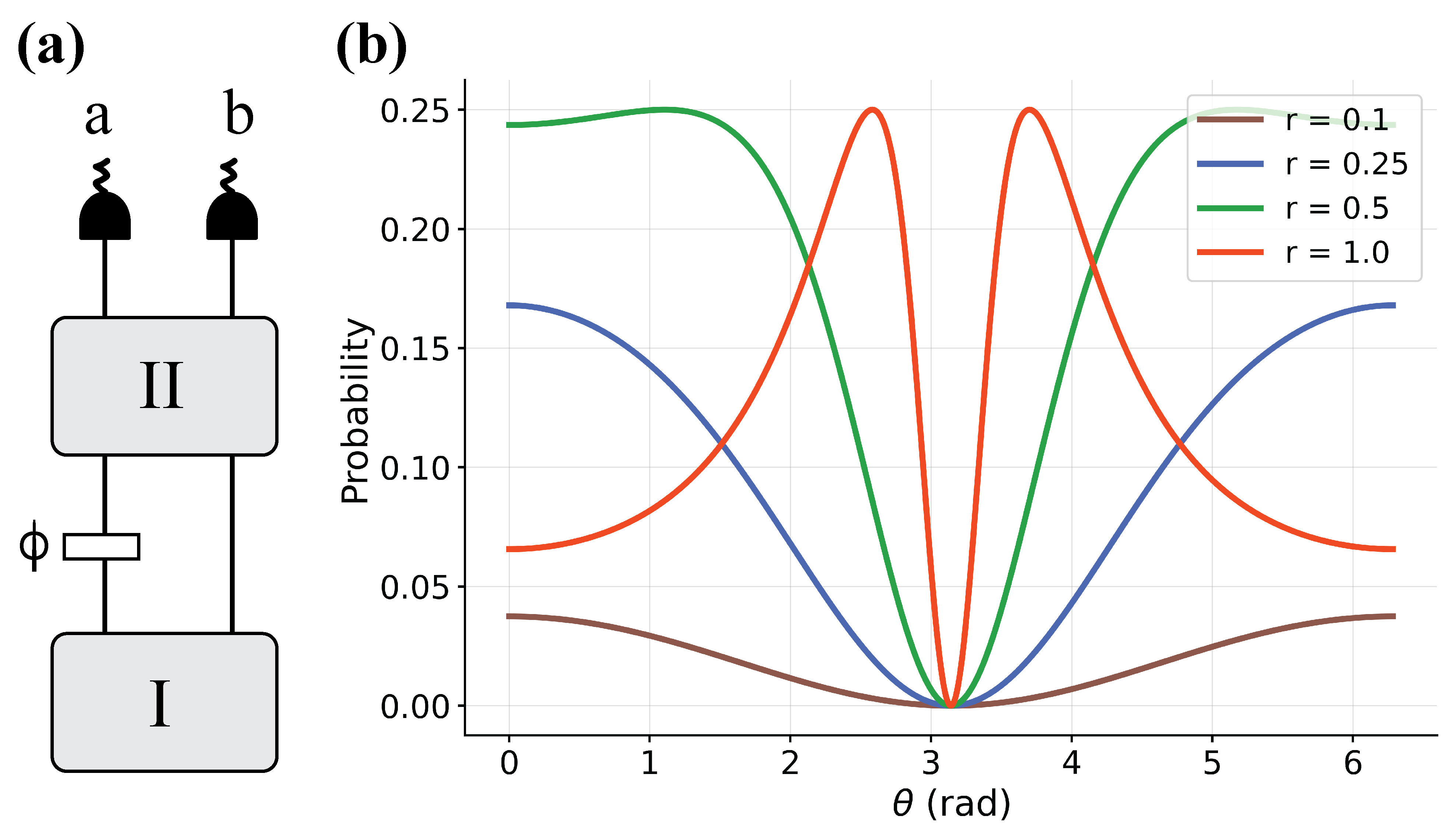}
    \caption{Two-crystal setup and two-photon coincidence probability. 
    \textbf{(a)} The experimental scheme shows a phase shifter placed between nonlinear crystals I and II, with detections in the output paths $a$ and $b$. \textbf{(b)} The two-photon coincidence probability $P_{\ket{1,1}}$ (one photon in each output) is plotted as a function of the relative phase for different squeezing strengths $r$, assuming $r_{1}=r_{2}=r$.}
    \label{2crystal}
\end{figure}

One can immediately see interesting physical effects by putting two crystals sequentially, in such a way that the paths of the photons produced in the first crystals are overlapped with the paths of the second crystal, see Fig.~\ref{2crystal} (a). This setup is also sometimes called $\mathrm{SU(1,1)}$ nonlinear interferometer \cite{yurke19862, chekhova2016nonlinear, ou2020quantum}. We now apply Eq.~\eqref{eq:S2Fockpq} sequentially to both crystals (the output modes of the first crystal will also act as input to the second one), the amplitude for observing one photon pair $\ket{1,1}$ in output modes is
\begin{align}
&A(r_1,r_2,\phi)_{\ket{1,1}}=\bra{1,1}S_{ab}(r_2)\Phi_{a}(\phi)S_{ab}(r_1)\ket{0,0}\nonumber\\
&=\sum_{n=0}^{\infty}
        \frac{(-e^{i\phi}\tanh r_{1}\tanh r_{2})^{n}}{\cosh r_{1}\, \cosh^3 r_{2}\,\tanh r_{2}}\Big(n-\sinh^2 r_{2}\Big).
\label{eq:herzog11full}
\end{align}
Here $r_1$ and $r_2$ are the squeezing parameters of nonlinear crystals~I and~II, respectively. $S_{ab}(r)$ is the two-mode squeezing operator acting on modes $a$ and $b$, and
$\Phi_a(\phi)=\exp(i\phi\,\hat n_a)$ is a phase-shift operator acting on mode $a$, with $\hat n_a=\hat a^\dagger\hat a$ the photon-number operator. In the Fock basis it acts as
$\Phi_a(\phi)\ket{n}_a = e^{in\phi}\ket{n}_a$.

For real, finite squeezing parameters (i.e., physically realizable values of $r_1$ and $r_2$), the series in Eq.~\eqref{eq:herzog11full} converges since $|-e^{i\phi}\tanh r_{1}\tanh r_{2}|<1$, leading to
\begin{equation}
A(r,\phi)_{\ket{1,1}}=-\frac{\left(\tanh r_2+e^{i\phi}\tanh r_1\right)\,\text{sech} r_1\,\text{sech} r_2}{\left(1+e^{i\phi}\tanh r_1 \tanh r_2\right)^{2}}.
\label{eq:herzog11r_ampr1r2}
\end{equation}
When both crystals are pumped with equal squeezing strengths ($r_1=r_2=r$), this expression simplifies to
\begin{align}
A(r,\phi)_{\ket{1,1}}=-\frac{\left(1+e^{i\phi}\right)\,\text{sech}^{2} r\,\tanh r}{\left(1+e^{i\phi}\tanh^{2} r\right)^{2}}.
\label{eq:herzog11r_amp}
\end{align}

In the low-gain regime, expanding the amplitude in Eq.~\eqref{eq:herzog11r_amp} with respect to $r$, we find
\begin{align}
A_{\ket{1,1}}(r_{\text{small}}\rightarrow 0 ,\phi)\approx -\left(1+e^{i\phi}\right)r+O(r^3)
\label{eq:herzog11rsmall}
\end{align}
and the corresponding probability of observing the state $\ket{1,1}$ is approximately
\begin{align}
P(r_\text{small},\phi)_{\ket{1,1}}=|A_{\ket{1,1}}(r_\text{small},\phi)|^2\approx4r^2\cos^2(\frac{\phi}{2})
\label{eq:herzog11r_P}
\end{align}
which exhibits a sinusoidal oscillation with the phase $\phi$.

This behavior reveals the phenomenon of \textit{Frustrated two-photon creation} \cite{herzog1994frustrated}, where one has a superposition of the photons being created in the first or the second crystal. When the relative phase between the two crystals is set to $\phi=\pi$, the two possibilities interfere destructively, completely canceling the photon pair creation. This phenomenon has been extensively studied in the context of quantum metrology \cite{hochrainer2022quantum, barreto2022quantum, bowen2023quantum}, especially as it allows for the two photons in paths $a$ and $b$ to have different frequencies, which can be exploited in practical applications \cite{kalashnikov2016infrared, kviatkovsky2020microscopy}.

\begin{figure*}
\centering
\includegraphics[width=1\linewidth]{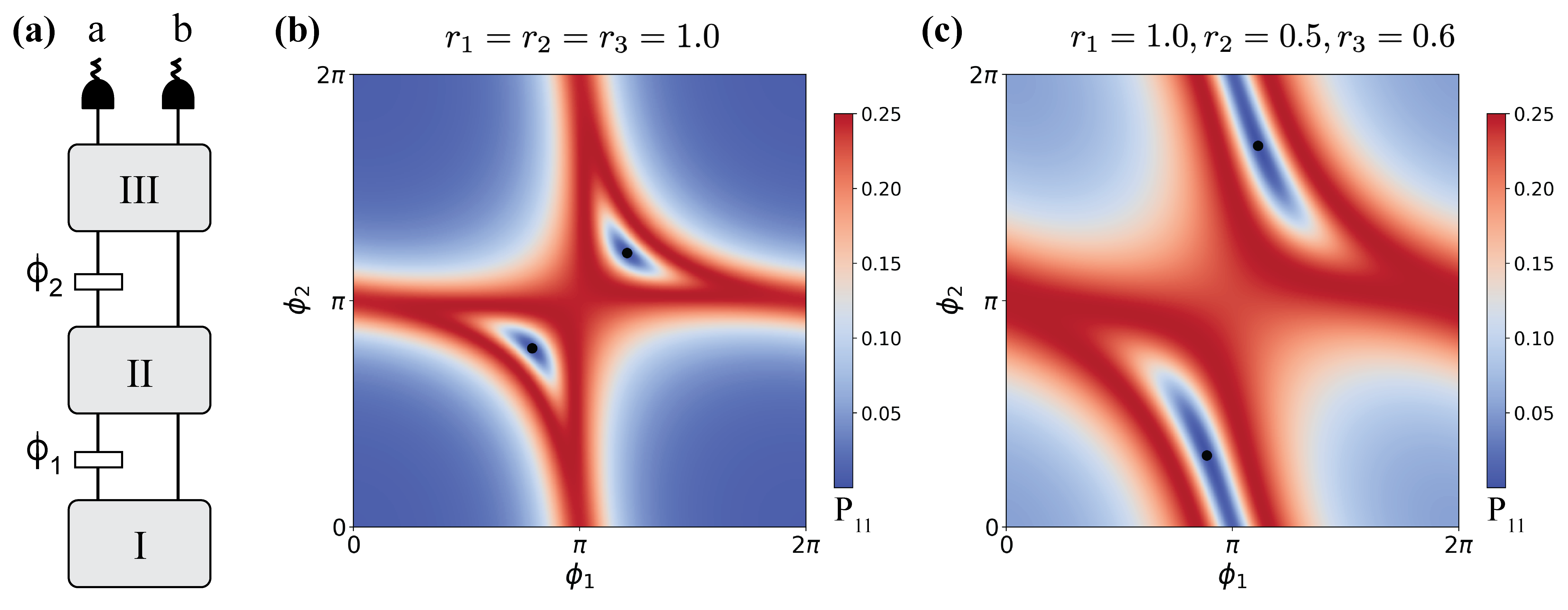}
\caption{Three-crystal setup and two-photon coincidence probability. \textbf{(a)} The experimental scheme includes two phase shifters placed between the three sequential nonlinear crystals. \textbf{(b)} The coincidence probability $P_{\ket{1,1}}$ for equal squeezing strengths $r_{1}=r_{2}=r_{3}=1.0$ shows a coincidence zero due to perfect destructive interference (black dots). \textbf{(c)} The coincidence probability for asymmetric pump powers ($r_{1}=1.0$, $r_{2}=0.5$, $r_{3}=0.6$) also exhibits destructive interference of photon pairs (black dots).}
\label{3crystal}
\end{figure*}
In the high-gain regime where $r$ is large, other interesting phenomena emerge. Conventionally, this regime is studied using the language of continuous-variable quantum mechanics ~\cite{braunstein2005quantum} and the application of homodyne measurements. This is practically (both experimentally and theoretically) useful~\cite{anisimov2010quantum,plick2010coherent,machado2020optical,santandrea2023lossy}, but makes it difficult to see the concrete behavior of the individual terms in the Fock basis.

Using the explicit Fock-basis expression Eq.~\eqref{eq:S2Fockpq}, we find that the simple sinusoidal oscillation of $P_{\ket{1,1}}(r,\phi)$ changes as $r$ increases: the probability of observing $\ket{1,1}$ becomes drastically changes and highly sensitive to small phase shifts near $\phi=\pi$. This sensitivity can be quantified by the second derivative (or the curvature) $C(r,\phi)$ of Eq.~\eqref{eq:herzog11r_P},
\begin{align}
C(r,\phi=\pi)
&=\left.\frac{\partial^2 P(r,\phi)_{\ket{1,1}}}{\partial \phi^2}\right|_{\phi=\pi}=\frac{\sinh^{2}(2r)}{2} %2,\cosh^{2}(r)\sinh^{2}(r)
\label{eq:herzogcurvature}
\end{align}

For large $r$, $\sinh(2r)\sim\tfrac{1}{2}e^{2r}$, so the curvature scales approximately as $C(r,\pi)\propto e^{4r}$. The phase response therefore grows exponentially with increasing squeezing strength, as also numerically shown in  Fig.~\ref{2crystal} (b). This is very useful for precise measurements of small phase changes~\cite{lindner2023high}, if the experimental challenges such as achieving large squeezing and photon-number-resolving detection can be overcome.

\subsection*{Two-photon interference among three crystals}
One interesting recent interference effect involves three sequential crystals ~\cite{jiang2025subjective} (described in more detail in \cite{Hochrainer2020PhD}), as shown in Fig.~\ref{3crystal}. When grouping the crystals into pairs, one can create a situation where a classical description would lead to two contradicting descriptions. We apply Eq.~\eqref{eq:S2Fockpq} and get the general amplitude description of the two-photon term $\ket{1,1}$ at the output
\begin{align}
A(&r_{1},r_{2},r_{3},\phi_1,\phi_2)_{\ket{1,1}}\nonumber\\
&= \bra{1,1}S_{ab}(r_3)\Phi_a(\phi_2)S_{ab}(r_2)\Phi_{a}(\phi_1)S_{ab}(r_1)\ket{0,0}\nonumber\\
&= -\,s_1 s_2 s_3\;
\frac{\,t_3\!\left(1+e^{i\phi_1} t_1 t_2\right)\;+\;e^{i\phi_2}\!\left(t_2+e^{i\phi_1} t_1\right)}
{\Big(1+e^{i\phi_1} t_1 t_2 + e^{i\phi_2} t_2 t_3 + e^{i(\phi_1+\phi_2)} t_1 t_3\Big)^{\!2}}
\label{eq:threecrystalFullForm}
\end{align}
%\begin{widetext}
%\begin{align}
%A(r_{1},r_{2},r_{3},\phi_1,\phi_2)_{\ket{1,1}}&= \bra{1,1}S_{ab}(r_3)\Phi_a(\phi_2)S_{ab}(r_2)\Phi_{a}(\phi_1)S_{ab}(r_1)\ket{0,0}\nonumber\\
%&= -\frac{\operatorname{sech} r_1 \operatorname{sech} r_2 \operatorname{sech} r_3 \Big[\tanh{r_3}\!\left(1+e^{i\phi_1} \tanh{r_1} \tanh{r_2}\right)\;+\;e^{i\phi_2}\!\left(\tanh{r_2}+e^{i\phi_1} \tanh{r_1}\right)\Big]}{\Big(1+e^{i\phi_1} \tanh{r_1} \tanh{r_2}+ e^{i\phi_2} \tanh{r_2}\tanh{r_3} + e^{i(\phi_1+\phi_2)} \tanh{r_1}\tanh{r_3}\Big)^{\!2}}
%\label{eq:threecrystalFullForm}
%\end{align}
%\end{widetext}
with $t_i=\tanh{r_i}$ and $s_i=\operatorname{sech} r_i$. The surprising behavior described in Ref.~\cite{jiang2025subjective} is given by
\begin{align}
A(r,r,r,\pi,\pi)_{\ket{1,1}}=-\operatorname{sech} r \tanh{r}, 
\end{align}
which leads to a situation where crystal I and II together (and independently from the rest) as well as crystal II and III together appear to exhibit perfect destructive interference, thus concluding that any of the remaining photons must be created only in the other crystal (crystal III and I, respectively). This conclusion is contradicting, which shows one cannot use such semi-classical description to assign definite creation events to specific crystals.

\begin{figure*}
    \centering
    \includegraphics[width=1\linewidth]{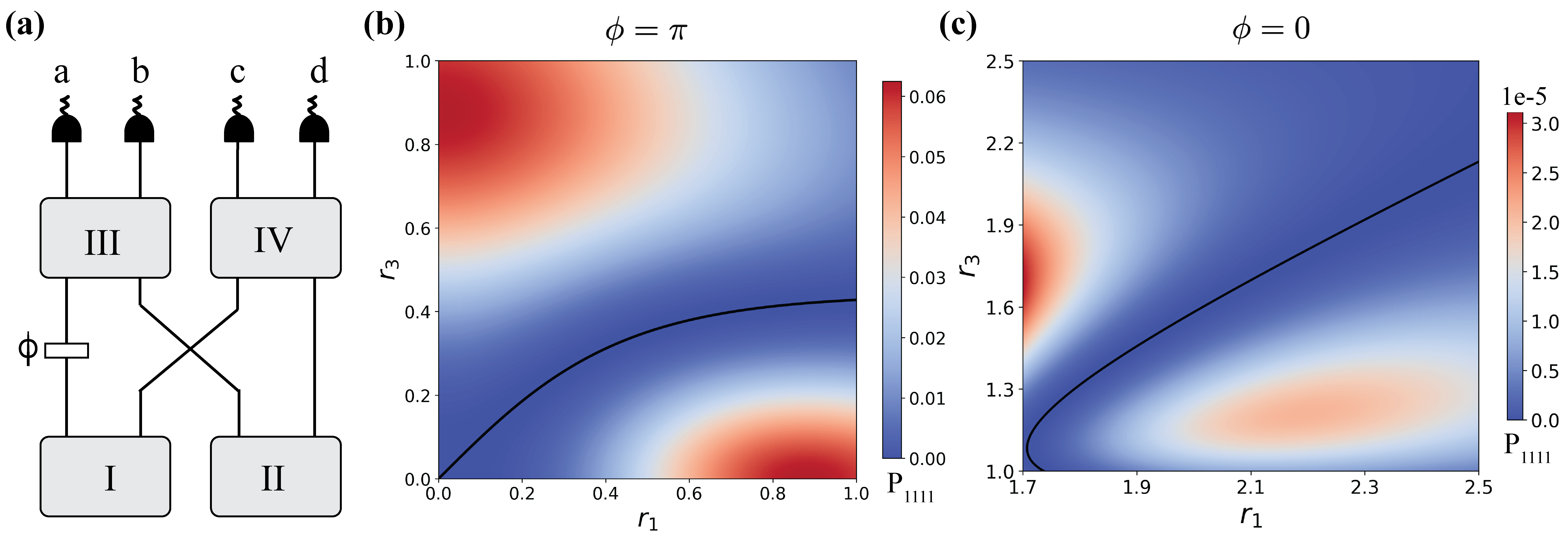}
\caption{Four-crystal setup and four-photon coincidence probability. \textbf{(a)} The experimental scheme includes a phase shifter placed between crystals I and III in path $a$. \textbf{(b)} The four-photon coincidence probability $P_{\ket{1,1,1,1}}$ is shown for phase $\phi=\pi$, with squeezing strengths set to $r_{1}=r_{2}$ and $r_{3}=r_{4}$. \textbf{(c)} The probability for phase $\phi=0$ is shown for high squeezing, with $r_{1}=r_{2}$ and $r_{4}=2r_{3}$.}
    \label{4crystal}
\end{figure*}
Analyzing Eq.~\eqref{eq:threecrystalFullForm} with $r_i=r$, we find destructive interference of $A(r,r,r,\phi_1,\phi_2)_{\ket{1,1}}=0$ only for
\begin{align}
\phi_1=\phi_2 \in \Bigl\{\vartheta(r),\ \ 2\pi-\vartheta(r))\Bigr\},
\label{threecrystalRRRzero}
\end{align}
with $\vartheta(r)=2\arccos\!\big(\tfrac{\operatorname{sech} r}{2}\big)$.

For the low-gain regime, the amplitude in Eq.~\eqref{eq:threecrystalFullForm} can be expanded as  
%\begin{align}
%A(r,r,r,\phi_1,\phi_2)_{\ket{1,1}}
  %\approx -\Bigl[1 + e^{i\phi_2}\bigl(1 + e^{i\phi_1}\bigr)\Bigr]\,r
  %+ O(r^{3}).
%\label{threecrystal}
%\end{align}
\begin{align}
A&(r, r,r,\phi_1,\phi_2)_{\ket{1,1}}\nonumber\\
  & \approx -\Bigl(1 + e^{i\phi_2}\bigl(1 + e^{i\phi_1}\bigr)\Bigr)r + \Bigr[2e^{2i\phi_2}(1+e^{i\phi_1})^2\nonumber\\
  &\quad + \frac{e^{i\phi_2}}{6}(1+e^{i\phi_1})(23+12e^{i\phi_1}) +\frac{11}{6} +e^{i\phi_1} \Bigr]r^{3} + O(r^{4}).
\label{threecrystal}
\end{align}

At the lowest order in $r$, the amplitude vanishes when  
$1 + e^{i\phi_2} + e^{i(\phi_1+\phi_2)} = 0$.  
This condition is satisfied for 
\begin{align}
    \phi_1 = \phi_2 = \vartheta(r) = \tfrac{2\pi}{3} (\text{or}\, \tfrac{4\pi}{3})\,\,\, \text{modulo}\,\, 2\pi,\nonumber
\end{align}
reflecting the well-known property that the three cube roots of unity sum to zero, $1 + e^{i\tfrac{2\pi}{3}} + e^{i\tfrac{4\pi}{3}} = 0$.  

If we use $\vartheta(r=\tfrac{2\pi}{3})$, higher-order terms remain, giving $A \sim r^{3}$ for small but finite $r$. A more accurate cancellation requires the phase to drift slightly with the gain parameter. Including the next correction, we find that the destructive-interference condition is satisfied when
\[
\vartheta(r) = \frac{2\pi}{3} - \frac{r^{2}}{\sqrt{3}} + O(r^{4})
\quad\text{or}\quad
\vartheta(r) = \frac{4\pi}{3} + \frac{r^{2}}{\sqrt{3}} + O(r^{4}),
\]
which suppresses the $r^{3}$ contribution. 

For high values of $r$, the phases need to be adjusted such that they can compensate for the extra contributions (Fig.~\ref{3crystal} (b)). We find a general analytical condition for cancellation of $\ket{1,1}$ term (as before, $t_i=\tanh{r_i}$ and $s_i=\operatorname{sech} r_i$)
\begin{align}
\phi_1 &\in \Bigl\{\,\arccos C,\ \ 2\pi-\arccos C\,\Bigr\}\quad (\text{mod }2\pi),
\\[3pt]
&\quad\text{where}\quad C = \frac{t_3^2\!\left(1+t_1^2 t_2^2\right)-t_1^2-t_2^2}{2\,t_1 t_2\,(1-t_3^2)}\nonumber
\end{align}
together with
\begin{align}
\phi_2\equiv \Arg\!\left(-\,t_3\,\frac{1+t_1 t_2 e^{i\phi_1}}{\,t_1 e^{i\phi_1}+t_2\,}\right)
\quad(\text{mod }2\pi).
\end{align}
under the constraint 
\[
\frac{|t_1 - t_2|}{1 - t_1 t_2} \ \le\ t_3 \ \le\ \frac{t_1 + t_2}{1 + t_1 t_2}.
\]
This result shows that, for any set of squeezing parameters, one can always find a pair of phases $(\phi_1,\phi_2)$ that removes the two-photon coincidence component $\ket{1,1}$.

\subsection*{Four-photon interference in a four-crystal configuration}
A more recent development is the theoretical prediction \cite{gu2019quantum} and experimental observation \cite{feng2023chip,qian2023multiphoton,wang2025violation} of a new multi-photon, nonlocal quantum interference effect. The simplest version consists of two rows of two crystals each, see Fig.~\ref{4crystal} (a).

Let's look at an interesting special case: The only way to get exactly one photon in each of the four detectors (i.e., seeing the four-photon state $\ket{1,1,1,1}$) is that either the lower row of crystals, i.e. crystals I and II each produce a pair, or the upper row with crystal III and IV produce a pair each. No other combination of firing pattern produces one photon in each output path. As in the previous case, if one can change the phase between the two crystals I and III, we can destructively (and constructively) interfere the two possible origins of the four-photon state. The amplitude of the $\ket{1,1,1,1}$ term can be found as
\begin{align}
&A(r_{1},r_{2},r_{3},r_{4},\phi)_{\ket{1,1,1,1}}\nonumber\\ 
&=\bra{1,1,1,1}S_{\text{cd}}(r_4)S_{\text{ab}}(r_3)\Phi_{a}(\phi)S_{\text{bd}}(r_2)S_{\text{ac}}(r_1)\ket{0,0,0,0}\nonumber\\
&=\frac{\operatorname{sech} r_1 \operatorname{sech} r_2 \operatorname{sech} r_3 \operatorname{sech} r_4}{2\left(e^{i\phi}\,\tanh r_{1}\,\tanh r_{2}\,\tanh r_{3}\,\tanh r_{4} -1\right)^{3}}\nonumber\\
&\quad\times\biggl(\frac{e^{i\phi}(\cosh 2r_3\cosh 2r_4 -3)\tanh r_{1}\,\tanh r_{2}}{\cosh^2 r_3 \cosh^2 r_4}\nonumber \\
&\quad -2(1+e^{2i\phi} \tanh^2 r_{1}\,\tanh^2 r_{2})\tanh r_{3}\,\tanh r_{4}\biggr). 
\label{eq:amp1111}
\end{align}

We can simplify for cases where all of the squeezing parameters are the same, $r_i \to r$, which leads to 
\begin{align}
A(&r,r,r,r, \phi)_{\ket{1,1,1,1}}\nonumber\\
  &= -\,\frac{\operatorname{sech}^{8} r\,\tanh^{2} r}
          {8\bigl(e^{i\phi}\tanh^{4} r -1\bigr)^{3}}\times\Bigl[3 + 4\cosh 2r + \cosh 4r\nonumber\\
  &\quad + 2e^{i\phi}\bigl(5-\cosh 4r\bigr)
        + 8e^{2 i\phi}\sinh^{4} r\Bigr].
\label{eq:amp1111SameR}
\end{align}

In the low-gain regime, for small values of $r$, we can expand to the 4th order in $r$ and find that the amplitude of the $\ket{1,1,1,1}$ term is 
\begin{align}
A(&r_{\text{small}},r_{\text{small}},r_{\text{small}},r_{\text{small}},\phi)_{\ket{1,1,1,1}}\nonumber\\
  &=\bigl(1 + e^{i\phi}\bigr) \, r^{2}
  \;-\; \frac{4}{3}\,\bigl(2 + 5 e^{i\phi}\bigr) \, r^{4}
  + \mathcal{O}\!\bigl(r^{6}\bigr).
\label{eq:amp1111SameRLowGain}  
\end{align}
The second-order term shows a complete cancellation of the term $\ket{1,1,1,1}$ when the phase is $\phi=\pi$. This case corresponds to the possibility that the four photons are either created in the lower two crystals or in the upper two crystals~\cite{gu2019quantum, qian2023multiphoton, feng2023chip}. Exactly the same as in the 2-photon 2-crystal case in Fig.~\ref{2crystal} (a), these two possibility cancel each other if the phase is $\phi=\pi$. 

Interestingly, however, in contrast to the case of 2-photon interference, the higher-order terms wash out the interference pattern (also see Fig.~\ref{4crystal} (b)). In fact, the analytical expression Eq.~\eqref{eq:amp1111SameR} shows explicitly that perfect destructive interference of the $\ket{1,1,1,1}$ term cannot occur once the squeezing parameter $r$ exceeds zero. This is because the amplitude in Eq.~\eqref{eq:amp1111SameR} could vanish only if the polynomial in $e^{i\phi}$ inside the brackets were zero, but this quadratic has no roots on the unit circle for any non-zero gain ($r>0$). Thus, the amplitude remains strictly non-zero beyond the low-gain regime.

However, we can find an interesting case of perfect cancellation if not all crystals have the same squeezing $r$, but crystals in each row have the same $r_i$, e.g., from Eq.~\eqref{eq:amp1111} we set $r_1=r_2$ and $r_3=r_4$. This gives
\begin{align}
A(&r_{1},r_{1}, r_{3},r_{3},\phi)_{\ket{1,1,1,1}}\nonumber\\
&= \frac{\operatorname{sech}^{2}r_{1}\,\operatorname{sech}^{2}r_{3}}
        {4\bigl(e^{i\phi}\tanh^{2}r_{1}\tanh^{2}r_{3}-1\bigr)^{3}}\times \nonumber\\
&\quad\Bigg[
        e^{i\phi}(-5+\cosh 4r_{3})\operatorname{sech}^{4}r_{3}\,\tanh^{2}r_{1}\nonumber\\
        &\quad\quad-\,4\,\bigl(1+e^{2i\phi}\tanh^{4}r_{1}\bigr)\tanh^{2}r_{3}
      \Bigg].
\label{eq:amp1111r1r3}
\end{align}
At $\phi=\pi$, this simplifies to
\begin{align}
A(&r_{1},r_{1}, r_{3},r_{3}, \pi)_{\ket{1,1,1,1}}\nonumber\\
&= \frac{%
        2\cosh^{2}2r_{1}\,\cosh4r_{3} - 3\cosh 4r_{1} + 1 %
      }{%
        2\bigl(\cosh2r_{1}\,\cosh2r_{3} + 1 \bigr)^{3}%
      }.
\label{eq:amp1111r1r3PhiPi}
\end{align}
Perfect destructive interference occurs when
\begin{align}
r_3 = \frac{\operatorname{arcsinh}(\tanh 2r_1)}{2},
\label{eq:r3SimplifiedZERO}
\end{align}
demonstrating that appropriate choice of asymmetric squeezing enables high-gain cancellation (Fig.~\ref{4crystal} (b)).

Interestingly, perfect destructive interference can also occur with zero relative phase when the upper row has asymmetric squeezing. Setting $r_1=r_2$, $r_3\not=r_4$ and $\phi=0$ in analytical expression Eq.~\eqref{eq:amp1111} gives
\begin{align}
A(&r_{1},r_{1}, r_{3},r_{4}, 0)_{\ket{1,1,1,1}}\nonumber\\
&= \frac{\operatorname{sech}^{6}r_{1} \operatorname{sech}^{3}r_{3}\operatorname{sech}^{3}r_{4}
      }{
        16\bigl(\tanh^{2}r_{1}\,\tanh r_{3}\,\tanh r_{4} -1 \bigr)^{3}
      }\nonumber\\
     &\quad\times\Bigg[ 2\bigl(\cosh{2r_{3}}\cosh{2r_{4}}-3\bigr)\sinh^{2}{2r_{1}}\nonumber\\
     &\quad-\bigl(3+\cosh{4r_{4}}\bigr)\sinh{2r_{3}}\sinh{2r_{4}}\Bigg]
\label{eq:amp1111r1r3r4Phi0}
\end{align}

At first glance, one might expect that with zero relative phase ($\phi=0$), the four-photon term $\ket{1,1,1,1}$ would always appear if one follows a naive low-gain intuition, where all crystals are pumped coherently and interfere constructively. However, when the two crystals in the upper row have unequal squeezing strengths ($r_3 \neq r_4$), a nontrivial destructive interference emerges. In this case, the amplitude vanishes when
\begin{align}
&r_1=\tfrac12\,\operatorname{arsinh}
\!\left(
\sqrt{\dfrac{2\,\sinh(2r_3)\,\sinh(2r_4)}{\cosh\!\bigl(2(r_3-r_4)\bigr)-3}}
\right)\, \nonumber\\
&|r_3-r_4|>\operatorname{arccosh}(1)\nonumber
\label{eq:amp1111r1r3r4Phi0_condidions}
\end{align}
The interference pattern is shown in Fig.~\ref{4crystal} (c).

An intuitive explanation for cancellation at $\phi=0$ emerges from recognizing that photon pair creation and annihilation carry a relative phase of $\pi$, as shown in Eq.~\eqref{eq:s2operator}. Any process involving creation followed by annihilation (or vice versa) returns to the original quantum state but with amplitude shifted by $\pi$. This is precisely the physics underlying the single-crystal two-photon interference in Eq.~\eqref{eq:singleCrystalInterference}. In the four-crystal geometry with asymmetric squeezing, complex combinations of creation and annihilation events can interfere destructively even when the geometric phase between rows is zero, provided the squeezing parameters are appropriately tuned.

\section*{Conclusion}
We present an exact representation of the two-mode squeezing operator acting on arbitrary Fock states. This analytical expression enables us to reinterpret known quantum-interference phenomena and to predict new effects in multi-crystal parametric down-conversion systems. Specifically, we have (i) described two-photon interference in single-crystal systems, (ii) analyzed two-photon interference in two-crystal configurations across both low- and high-gain regimes, (iii) identified the conditions for perfect destructive interference in three-crystal arrangements, and (iv) discovered new four-photon interference effects, including cases where perfect cancellation occurs only under asymmetric squeezing conditions.

While our analysis focuses on the lowest-order Fock components (e.g., $\ket{1,1}$ or $\ket{1,1,1,1}$), the derived expressions remain valid for arbitrary photon numbers. A systematic exploration of interference effects among higher-order Fock states may therefore lead to additional, nontrivial multi-photon interference phenomena and interesting photon-number states \cite{xue2025photon}. Beyond the few-source settings considered here, the exact analytic Fock-basis framework naturally extends to large-scale quantum-network architectures on integrated photonic platforms, where network-level interference could play a central role \cite{bao2023very}. It could also provide analytic guidance for quantum state generation, including high-dimensional and multipartite structures, complementing recent demonstrations and proposals \cite{hu2025observation, bernecker2025quantum}. The analytical treatment may further help identify robust implementations e.g., where phase fluctuations are limited.

Moreover, our analysis assumes a classical pump, whereas a general description would incorporate the quantum nature of the pump-depletion in parametric down-conversion, described by the three-mode squeezing operator $U=\exp \left(\zeta^* p^\dagger a b  - \zeta p a^\dagger b^\dagger\right)$, where $p$ represents the pump mode. This generalized model would allow a systematic analysis of quantum interference in experiments such as \cite{kopf2025conservation}. It would also be interesting in the future to investigate systems with probabilistic photon sources beyond pair-sources, such as genuine three-photon sources or N-photon sources~\cite{douady2004experimental, chekhova2005spectral, hammer2018dispersion}.

Finally, the analytical expression we demonstrate might be useful as the basis for classical simulators of quantum optics experiments, especially for the AI-driven discovery of new experiments with interesting properties that go beyond low-order approximations \cite{krenn2016automated,ruiz2023digital}.

\section*{Acknowledgments}
The authors thank Soeren Arlt, Xiaosong Ma and Kai Wang for many useful discussions. MK acknowledges support by the European Research Council (ERC) under the European Union’s Horizon Europe research and innovation programme (ERC-2024-STG, 101165179, ArtDisQ) and from the German Research Foundation DFG (EXC 2064/1, Project 390727645). X.G. acknowledges the support from the German Research Foundation DFG via SFB 1375 (NOA) and the Alexander von Humboldt Foundation.

\bibliography{ref}

@article{truax1985baker,
  title={Baker-Campbell-Hausdorff relations and unitarity of SU (2) and SU (1, 1) squeeze operators},
  author={Truax, D Rodney},
  journal={Physical Review D},
  volume={31},
  number={8},
  pages={1988},
  year={1985},
  publisher={APS},
  url={https://doi.org/10.1103/PhysRevD.31.1988}
}

@article{chiribella2006applications,
  title={Applications of the group SU (1, 1) for quantum computation and tomography},
  author={Chiribella, Giulio and D’Ariano, Giacomo M and Perinotti, Paolo},
  journal={Laser physics},
  volume={16},
  number={11},
  pages={1572--1581},
  year={2006},
  publisher={Springer},
  url={https://doi.org/10.1134/S1054660X06110119}
}

@book{kok2010introduction,
  title={Introduction to optical quantum information processing},
  author={Kok, Pieter and Lovett, Brendon W},
  year={2010},
  publisher={Cambridge university press},
  url={https://doi.org/10.1017/CBO9781139193658}
}

@article{chizhov1993photon,
  title={Photon statistics and phase properties of two-mode squeezed number states},
  author={Chizhov, AV and Murzakhmetov, BK},
  journal={Physics Letters A},
  volume={176},
  number={1-2},
  pages={33--40},
  year={1993},
  publisher={Elsevier},
  url={https://doi.org/10.1016/0375-9601(93)90312-N}
}

@article{gu2019quantum,
  title={Quantum experiments and graphs ii: Quantum interference, computation, and state generation},
  author={Gu, Xuemei and Erhard, Manuel and Zeilinger, Anton and Krenn, Mario},
  journal={Proceedings of the National Academy of Sciences},
  volume={116},
  number={10},
  pages={4147--4155},
  year={2019},
  publisher={National Academy of Sciences},
url={https://doi.org/10.1073/pnas.1815884116}
}

@article{cerf2020two,
  title={Two-boson quantum interference in time},
  author={Cerf, Nicolas J and Jabbour, Michael G},
  journal={Proceedings of the National Academy of Sciences},
  volume={117},
  number={52},
  pages={33107--33116},
  year={2020},
  publisher={National Academy of Sciences},
  url={https://doi.org/10.1073/pnas.2010827117}
}

@article{chen2025two,
  title={Two-particle quantum interference in a nonlinear optical medium: a witness of timelike indistinguishability},
  author={Chao Chen and Shu-Tian Xue and Yu-Peng Shi and Jing Wang and Zi-Mo Cheng and Pei Wan and Zhi-Cheng Ren and Michael G. Jabbour and Nicolas J. Cerf and Xi-Lin Wang and Hui-Tian Wang},
  journal={arXiv:2502.01480},
  year={2025},
  url={https://arxiv.org/abs/2502.01480},
}

@article{hu2025observation,
  title={Observation of Genuine High-dimensional Multi-partite Non-locality in Entangled Photon States},
  author={Hu, Xiao-Min and Huang, Cen-Xiao and d'Alessandro, Nicola and Cobucci, Gabriele and Zhang, Chao and Guo, Yu and Huang, Yun-Feng and Li, Chuan-Feng and Guo, Guang-Can and Gao, Xiaoqin and Huber, Marcus and Tavakoli, Armin and Liu, Bi-Heng},
  journal={Nature Communications},
  volume={16},
  number={1},
  pages={5017},
  year={2025},
  publisher={Nature Publishing Group UK London},
  url={https://doi.org/10.1038/s41467-025-59717-y}
}

@article{bao2023very,
  title={Very-large-scale integrated quantum graph photonics},
  author={Bao, Jueming and Fu, Zhaorong and Pramanik, Tanumoy and Mao, Jun and Chi, Yulin and Cao, Yingkang and Zhai, Chonghao and Mao, Yifei and Dai, Tianxiang and Chen, Xiaojiong and others},
  journal={Nature Photonics},
  volume={17},
  number={7},
  pages={573--581},
  year={2023},
  publisher={Nature Publishing Group UK London},
 url={https://doi.org/10.1038/s41566-023-01187-z}
}

@article{wang2025violation,
  title={Violation of Bell inequality with unentangled photons},
  author={Wang, Kai and Hou, Zhaohua and Qian, Kaiyi and Chen, Leizhen and Krenn, Mario and Aspelmeyer, Markus and Zeilinger, Anton and Zhu, Shining and Ma, Xiao-Song},
  journal={Science Advances},
  volume={11},
  number={31},
  pages={eadr1794},
  year={2025},
  publisher={American Association for the Advancement of Science},
  url={https://doi.org/10.1126/sciadv.adr1794}
}

@article{qian2023multiphoton,
  title={Multiphoton non-local quantum interference controlled by an undetected photon},
  author={Qian, Kaiyi and Wang, Kai and Chen, Leizhen and Hou, Zhaohua and Krenn, Mario and Zhu, Shining and Ma, Xiao-song},
  journal={Nature Communications},
  volume={14},
  number={1},
  pages={1480},
  year={2023},
  publisher={Nature Publishing Group UK London},
  url={https://doi.org/10.1038/s41467-023-37228-y}
}

@article{feng2023chip,
  title={On-chip quantum interference between the origins of a multi-photon state},
  author={Feng, Lan-Tian and Zhang, Ming and Liu, Di and Cheng, Yu-Jie and Guo, Guo-Ping and Dai, Dao-Xin and Guo, Guang-Can and Krenn, Mario and Ren, Xi-Feng},
  journal={Optica},
  volume={10},
  number={1},
  pages={105--109},
  year={2023},
  publisher={Optica Publishing Group},
  url={https://doi.org/10.1364/OPTICA.474750}
}

@article{dasgupta1996disentanglement,
  title={Disentanglement formulas: An alternative derivation and some applications to squeezed coherent states},
  author={DasGupta, Ananda},
  journal={American Journal of Physics},
  volume={64},
  number={11},
  pages={1422--1427},
  url={https://doi.org/10.1119/1.18183},
  year={1996}
}

@book{barnett2002methods,
  title={Methods in theoretical quantum optics},
  author={Barnett, Stephen and Radmore, Paul M},
  volume={15},
  year={2002},
  publisher={Oxford University Press},
url={https://doi.org/10.1093/acprof:oso/9780198563617.001.0001}
}

@article{santandrea2023lossy,
  title={Lossy SU (1, 1) interferometers in the single-photon-pair regime},
  author={Santandrea, Matteo and Luo, Kai-Hong and Stefszky, Michael and Sperling, Jan and Herrmann, Harald and Brecht, Benjamin and Silberhorn, Christine},
  journal={Quantum Science and Technology},
  volume={8},
  number={2},
  pages={025020},
  year={2023},
  publisher={IOP Publishing},
url={https://doi.org/10.1088/2058-9565/acc205}
}

@article{braunstein2005quantum,
  title={Quantum information with continuous variables},
  author={Braunstein, Samuel L and Van Loock, Peter},
  journal={Reviews of modern physics},
  volume={77},
  number={2},
  pages={513--577},
  year={2005},
  publisher={APS},
url={https://doi.org/10.1103/RevModPhys.77.513}
}

@article{anisimov2010quantum,
  title={Quantum Metrology with Two-Mode Squeezed Vacuum: Parity Detection Beats the Heisenberg Limit},
  author={Anisimov, Petr M and Raterman, Gretchen M and Chiruvelli, Aravind and Plick, William N and Huver, Sean D and Lee, Hwang and Dowling, Jonathan P},
  journal={Physical Review Letters},
  volume={104},
  number={10},
  pages={103602},
  year={2010},
  publisher={APS},
url={https://doi.org/10.1103/PhysRevLett.104.103602}
}

@article{bowen2023quantum,
  title={Quantum light microscopy},
  author={Bowen, WP and Chrzanowski, Helen M and Oron, Dan and Ramelow, Sven and Tabakaev, Dmitry and Terrasson, Alex and Thew, Rob},
  journal={Contemporary Physics},
  volume={64},
  number={3},
  pages={169--193},
  year={2023},
  publisher={Taylor \& Francis},
url={https://doi.org/10.1080/00107514.2023.2292380}
}

@article{kalashnikov2016infrared,
  title={Infrared spectroscopy with visible light},
  author={Kalashnikov, Dmitry A and Paterova, Anna V and Kulik, Sergei P and Krivitsky, Leonid A},
  journal={Nature Photonics},
  volume={10},
  number={2},
  pages={98--101},
  year={2016},
  publisher={Nature Publishing Group UK London},
url={https://doi.org/10.1038/nphoton.2015.252}
}

@article{ou2020quantum,
  title={Quantum SU (1, 1) interferometers: Basic principles and applications},
  author={Ou, Z\_Y and Li, Xiaoying},
  journal={APL Photonics},
  volume={5},
  number={8},
  year={2020},
  publisher={AIP Publishing},
  url={https://doi.org/10.1063/5.0004873}
}

@article{spagnolo2023non,
  title={Non-linear boson sampling},
  author={Spagnolo, Nicol{\`o} and Brod, Daniel J and Galv{\~a}o, Ernesto F and Sciarrino, Fabio},
  journal={npj Quantum Information},
  volume={9},
  number={1},
  pages={3},
  year={2023},
  publisher={Nature Publishing Group UK London},
url={https://doi.org/10.1038/s41534-023-00676-x}
}

@article{machado2020optical,
  title={Optical coherence tomography with a nonlinear interferometer in the high parametric gain regime},
  author={Machado, Gerard J and Frascella, Gaetano and Torres, Juan P and Chekhova, Maria V},
  journal={Applied Physics Letters},
  volume={117},
  number={9},
  year={2020},
  publisher={AIP Publishing},
  url={https://doi.org/10.1063/5.0016259}
}

@article{plick2010coherent,
  title={Coherent-light-boosted, sub-shot noise, quantum interferometry},
  author={Plick, William N and Dowling, Jonathan P and Agarwal, Girish S},
  journal={New Journal of Physics},
  volume={12},
  number={8},
  pages={083014},
  year={2010},
  publisher={IOP Publishing},
url={https://doi.org/10.1088/1367-2630/12/8/083014}
}

@article{ruiz2023digital,
  title={Digital discovery of 100 diverse quantum experiments with PyTheus},
  author={Ruiz-Gonzalez, Carlos and Arlt, S{\"o}ren and Petermann, Jan and Sayyad, Sharareh and Jaouni, Tareq and Karimi, Ebrahim and Tischler, Nora and Gu, Xuemei and Krenn, Mario},
  journal={Quantum},
  volume={7},
  pages={1204},
  year={2023},
  publisher={Verein zur F{\"o}rderung des Open Access Publizierens in den Quantenwissenschaften},
  url={https://doi.org/10.22331/q-2023-12-12-1204}
}

@article{barreto2022quantum,
  title={Quantum imaging and metrology with undetected photons: tutorial},
  author={Barreto Lemos, Gabriela and Lahiri, Mayukh and Ramelow, Sven and Lapkiewicz, Radek and Plick, William N},
  journal={Journal of the Optical Society of America B},
  volume={39},
  number={8},
  pages={2200--2228},
  year={2022},
  publisher={Optica Publishing Group},
url={https://doi.org/10.1364/JOSAB.456778}
}

@article{hochrainer2022quantum,
  title={Quantum indistinguishability by path identity and with undetected photons},
  author={Hochrainer, Armin and Lahiri, Mayukh and Erhard, Manuel and Krenn, Mario and Zeilinger, Anton},
  journal={Reviews of Modern Physics},
  volume={94},
  number={2},
  pages={025007},
  year={2022},
  publisher={APS},
url={https://doi.org/10.1103/RevModPhys.94.025007}
}

@article{kviatkovsky2020microscopy,
  title={Microscopy with undetected photons in the mid-infrared},
  author={Kviatkovsky, Inna and Chrzanowski, Helen M and Avery, Ellen G and Bartolomaeus, Hendrik and Ramelow, Sven},
  journal={Science Advances},
  volume={6},
  number={42},
  pages={eabd0264},
  year={2020},
  publisher={American Association for the Advancement of Science},
url={https://doi.org/10.1126/sciadv.abd0264}
}

@article{chekhova2016nonlinear,
  title={Nonlinear interferometers in quantum optics},
  author={Chekhova, MV and Ou, ZY},
  journal={Advances in Optics and Photonics},
  volume={8},
  number={1},
  pages={104--155},
  year={2016},
  publisher={Optical Society of America},
url={https://doi.org/10.1364/AOP.8.000104}
}

@article{yurke19862,
  title={SU (2) and SU (1, 1) interferometers},
  author={Yurke, Bernard and McCall, Samuel L and Klauder, John R},
  journal={Physical Review A},
  volume={33},
  number={6},
  pages={4033},
  year={1986},
  publisher={APS},
url={https://doi.org/10.1103/PhysRevA.33.4033}
}

@article{zheng2025nonlinear,
  title={Nonlinear-linear duality for multipath quantum interference},
  author={Zheng, Yi and Xu, Jin-Shi and Li, Chuan-Feng and Guo, Guang-Can},
  journal={Physical Review A},
  volume={112},
  number={3},
  pages={033705},
  year={2025},
  publisher={APS},
  url={https://doi.org/10.1103/6zhj-5t26}
}

@article{herzog1994frustrated,
  title={Frustrated two-photon creation via interference},
  author={Herzog, TJ and Rarity, JG and Weinfurter, H and Zeilinger, A},
  journal={Physical Review Letters},
  volume={72},
  number={5},
  pages={629},
  year={1994},
  publisher={APS},
url={https://doi.org/10.1103/PhysRevLett.72.629}
}

@article{hu2008two,
  title={Two-mode squeezed number state as a two-variable Hermite-polynomial excitation on the squeezed vacuum},
  author={Hu, Li-Yun and Fan, Hong-Yi},
  journal={Journal of Modern Optics},
  volume={55},
  number={13},
  pages={2011--2024},
  year={2008},
  publisher={Taylor \& Francis},
url={https://doi.org/10.1080/09500340801947629}
}

@article{jiang2025subjective,
  title={Subjective nature of path information in quantum mechanics},
  author={Jiang, Xinhe and Hochrainer, Armin and Kysela, Jaroslav and Erhard, Manuel and Gu, Xuemei and Yu, Ya and Zeilinger, Anton},
  journal={arXiv:2505.05930},
  year={2025},
  url={https://arxiv.org/abs/2505.05930},
}

@phdthesis{Hochrainer2020PhD,
  author    = {Armin Hochrainer},
  title     = {Path indistinguishability in photon pair emission},
  school    = {University of Vienna},
  address   = {Vienna, Austria},
  year      = {2020},
  doi       = {10.25365/thesis.64879},
  url       = {https://phaidra.univie.ac.at/o:1393189},
  note      = {Supervisor: Anton Zeilinger}
}

@article{krenn2016automated,
  title={Automated search for new quantum experiments},
  author={Krenn, Mario and Malik, Mehul and Fickler, Robert and Lapkiewicz, Radek and Zeilinger, Anton},
  journal={Physical Review Letters},
  volume={116},
  number={9},
  pages={090405},
  year={2016},
  publisher={APS},
url={https://doi.org/10.1103/PhysRevLett.116.090405}
}

@article{kopf2025conservation,
  title={Conservation of angular momentum on a single-photon level},
  author={Kopf, Lea and Barros, Rafael and Prabhakar, Shashi and Giese, Enno and Fickler, Robert},
  journal={Physical Review Letters},
  volume={134},
  number={20},
  pages={203601},
  year={2025},
  publisher={APS},
  url={https://doi.org/10.1103/PhysRevLett.134.203601}
}

@article{hammer2018dispersion,
  title={Dispersion tuning in sub-micron tapers for third-harmonic and photon triplet generation},
  author={Hammer, Jonas and Cavanna, Andrea and Pennetta, Riccardo and Chekhova, Maria V and Russell, Philip St J and Joly, Nicolas Y},
  journal={Optics Letters},
  volume={43},
  number={10},
  pages={2320--2323},
  year={2018},
  publisher={Optical Society of America},
  url={https://doi.org/10.1364/OL.43.002320}
}

@article{chekhova2005spectral,
  title={Spectral properties of three-photon entangled states generated via three-photon parametric down-conversion in a $\chi$ (3) medium},
  author={Chekhova, MV and Ivanova, OA and Berardi, V and Garuccio, Augusto},
  journal={Physical Review A},
  volume={72},
  number={2},
  pages={023818},
  year={2005},
  publisher={APS},
  url={https://doi.org/10.1103/PhysRevA.72.023818}
}

@article{douady2004experimental,
  title={Experimental demonstration of a pure third-order optical parametric downconversion process},
  author={Douady, J and Boulanger, B},
  journal={Optics letters},
  volume={29},
  number={23},
  pages={2794--2796},
  year={2004},
  publisher={Optical Society of America},
  url={https://doi.org/10.1364/OL.29.002794}
}

@article{bernecker2025quantum,
  title={Engineering of maximally entangled orbital angular momentum states via path identity},
  author={Bernecker, Richard and Baghdasaryan, Baghdasar and Fritzsche, Stephan},
  journal={Physical Review A},
  volume={112},
  number={6},
  pages={063701},
  year={2025},
  publisher={APS},
  url = {https://link.aps.org/doi/10.1103/9qrm-chgg}
}

@article{lindner2023high,
  title={High-sensitivity quantum sensing with pump-enhanced spontaneous parametric down-conversion},
  author={Lindner, Chiara and Kunz, Jachin and Herr, Simon J and Kie{\ss}ling, Jens and Wolf, Sebastian and K{\"u}hnemann, Frank},
  journal={APL photonics},
  volume={8},
  number={5},
  year={2023},
  publisher={AIP Publishing},
  url={https://doi.org/10.1063/5.0146025}
}

@article{jabbour2021multiparticle,
  title={Multiparticle quantum interference in Bogoliubov bosonic transformations},
  author={Jabbour, Michael G and Cerf, Nicolas J},
  journal={Physical Review Research},
  volume={3},
  number={4},
  pages={043065},
  year={2021},
  publisher={APS},
  url={https://doi.org/10.1103/PhysRevResearch.3.043065}
}

@article{miatto2020fast,
  title={Fast optimization of parametrized quantum optical circuits},
  author={Miatto, Filippo M and Quesada, Nicol{\'a}s},
  journal={Quantum},
  volume={4},
  pages={366},
  year={2020},
  publisher={Verein zur F{\"o}rderung des Open Access Publizierens in den Quantenwissenschaften},
  url={https://doi.org/10.22331/q-2020-11-30-366}
}

@article{xue2025photon,
  title={Photon-number entanglement from cascaded parametric down-conversion},
  author={Xue, Shu-Tian and Jiang, He and Wang, Jing and Ren, Zhi-Cheng and Wang, Xi-Lin and Wang, Hui-Tian},
  journal={Physical Review A},
  volume={112},
  number={6},
  pages={063719},
  year={2025},
  publisher={APS}
}

\clearpage
\onecolumngrid
\section*{Supplementary Information}
\subsection*{Normal Ordering Squeezing Operators}
The two-mode squeezing transformation, which models the action of a parametric down-conversion crystal, can be described by the operator
\begin{equation}
S_2(\zeta) = \exp \left(\zeta^* a b  - \zeta a^\dagger b^\dagger\right),\label{eq:s2operatorSI}
\end{equation}
where $a$ ($a^\dagger$) and $b$ ($b^\dagger$) denote the annihilation (creation) operators of the two modes, and $\zeta = r e^{i\theta}$ is the complex squeezing parameter. This transformation belongs to the Lie group $\mathrm{SU}(1,1)$, whose generators in the Schwinger representation are defined as ~\cite{chiribella2006applications}
\begin{equation}
    K_+ = a^\dagger b^\dagger, \,\,
    K_- = a b,\,\,
    K_0 = \frac{1}{2}\left( a^\dagger a + b^\dagger b + 1 \right),\label{eq:Koperators}
\end{equation}
which satisfy the following commutation relations:
\begin{align}
[K_0, K_\pm] = \pm K_\pm, \,\ [K_+, K_-] = -2K_0. \label{eq:commutation} 
\end{align}

In the following, we show how one can get the normal ordering expression of Eq.~\eqref{eq:S2normal_order} in the main text from Eq.~\eqref{eq:s2operatorSI}. Let's look at the transformation of the exponential operator \cite{dasgupta1996disentanglement}
\begin{align}
e^{\tau K_+ - \tau^* K_-}=e^{\beta K_+} e^{\gamma K_0} e^{\delta K_-} \, . \label{eq:S2normalorder}
\end{align}

Substituting $K_+$, $K_-$ from Eq.~\eqref{eq:Koperators} and setting $\tau = -r e^{i\theta}$ (i.e., $\tau^* = -r e^{-i\theta}$) in the left side of Eq.~\eqref{eq:S2normalorder}, we obtain
\begin{align}
e^{\tau K_+ - \tau^* K_-} \nonumber &= e^{-\zeta K_+ + \zeta^* K_-}= e^{-\zeta a^\dagger b^\dagger + \zeta^* ab}=\exp \left(\zeta^* a b  - \zeta a^\dagger b^\dagger\right)=S_2(\zeta)\label{eq:s2tonormalSI}
\end{align}
which identifies the two-mode squeezing operator $S_2(\zeta)$ in Eq.~\eqref{eq:s2operatorSI}. The goal now is to determine the coefficients $\beta$, $\gamma$, and $\delta$ in the right side of Eq.~\eqref{eq:S2normalorder}.

We begin by examining the operator on the left-hand side of Eq.~\eqref{eq:S2normalorder} and then define
\begin{equation}
    P={\tau K_+ - \tau^* K_-} \label{eq:operatorP}.
\end{equation}
leading to $S_2(\zeta)=e^{P}$. Now we compute $e^{P}K_0 e^{-P}$, which can be expanded using the Baker–Campbell–Hausdorff expansion as \cite{dasgupta1996disentanglement}
\begin{equation}
      e^{P}K_0 e^{-P}=\sum_{n=0}^\infty \frac{[P, K_0]_{n}}{n!}.
      \label{eq:BCHLemma}
 \end{equation}
The nested commutators are recursively defined as
\begin{equation}
 [P, K_0]_{n}=[P, [P, K_0]_{n-1}],\, \text{with}\quad [P, K_0]_{0}=K_0.
 \label{eq:BCHFcommutators}
\end{equation}

As we set the parameter $\tau = -\zeta = -r e^{i\theta} = r e^{i(\theta+\pi)}= \lambda e^{i\phi} $, we identify $\lambda=|\tau |=r$, $\phi=\text{Arg}(\tau)=\theta+\pi$. Using Eq.~\eqref{eq:BCHFcommutators}, we calculate the first few commutators explicitly:
\begin{align}
[P, K_0]_{1}&=[P, [P, K_0]_{0}]=[\tau K_+ - \tau^* K_-, K_0]=\tau [K_+, K_0]-\tau^*[K_-, K_0]= -\tau K_+ - \tau^* K_- -2\lambda \frac{(e^{i\phi}K_+ + e^{-i\phi}K_-)}{2},\nonumber\\
[P, K_0]_{2}&=[P, [P, K_0]_{1}]=[\tau K_+ - \tau^* K_-, -\tau K_+ - \tau^* K_-]= -\tau\tau^* [K_+, K_-]+\tau\tau^*[K_-, K_+]= (2\lambda)^{2}K_0,\nonumber\\
[P, K_0]_{3}&=[P, [P, K_0]_{2}]=[\tau K_+ - \tau^* K_-, (2\lambda)^{2}K_0]= (2\lambda)^{2}[\tau K_+ - \tau^* K_-, K_0]=-(2\lambda)^{3} \frac{(e^{i\phi}K_+ + e^{-i\phi}K_-)}{2}. \nonumber 
\end{align}
One can continue and obtain the following general result:
\begin{align}
&[P, K_0]_{2n}= (2\lambda)^{2n}K_0\, ,\,\,\, [P, K_0]_{2n+1}= - (2\lambda)^{2n+1}\frac{(e^{i\phi}K_+ + e^{-i\phi}K_-)}{2}
\end{align} 

Therefore, we can express $e^{P}K_0 e^{-P}$ as
\begin{align}
e^{P}K_0 e^{-P}&=\sum_{k=2n}^\infty \frac{[P, K_0]_{2n}}{(2n)!} + \sum_{k=2n+1}^\infty \frac{[P, K_0]_{2n+1}}{(2n+1)!}\label{eq:BCHLemmaFull}
\end{align}

With the hyperbolic functions, 
\begin{align}
\cosh x=1+ \frac{x^2}{2!}+  \frac{x^4}{4!} +...=\sum_{n=0}^\infty \frac{x^{2n}}{(2n)!}\, , \,\,\, \sinh x=1+ \frac{x^3}{3!}+  \frac{x^5}{5!} +...=\sum_{n=0}^\infty \frac{x^{2n+1}}{(2n+1)!}
\end{align}
one can rewrite Eq.~\eqref{eq:BCHLemmaFull} as
\begin{align}
e^{P}K_0 e^{-P}&=\sum_{k=2n}^\infty \frac{[P, K_0]_{2n}}{(2n)!} + \sum_{k=2n+1}^\infty \frac{[P, K_0]_{2n+1}}{(2n+1)!}=\sum_{k=2n}^\infty \frac{(2\lambda)^{2n}K_0}{(2n)!} + \sum_{k=2n+1}^\infty \frac{- (2\lambda)^{2n+1}(e^{i\phi}K_+ + e^{-i\phi}K_-)}{2\times(2n+1)!}\nonumber\\
&=K_0\cosh(2\lambda)-\frac{\sinh(2\lambda)(e^{i\phi} K_+ + e^{-i\phi}K_-)}{2}\label{eq:k0}
\end{align}

Similarly, one can do the same for $K_- $ and $K_-$ with $e^{P}(\cdot)e^{-P}$, leading to
\begin{align}
&e^{P}K_+ e^{-P}=-e^{-i\phi}\sinh(2\lambda)K_0+ \cosh^{2}(\lambda)K_+ + e^{-2i\phi}\sinh^{2}(\lambda)K_-\label{eq:kp}\\
&e^{P}K_- e^{-P}=-e^{i\phi}\sinh(2\lambda)K_0+ \cosh^{2}(\lambda)K_- + e^{2i\phi}\sinh^{2}(\lambda)K_+\label{eq:km}
\end{align}

We now examine the operator on the right-hand side of Eq.~\eqref{eq:S2normalorder} and calculate
\begin{align}
VK_0 V^{-1}=(e^{\beta K_+} e^{\gamma K_0} e^{\delta K_-})K_0 (e^{-\delta K_-} e^{-\gamma K_0} e^{-\beta K_+})\label{eq:Vk0V}, \text{where  } V=e^{\beta K_+} e^{\gamma K_0} e^{\delta K_-}.
\end{align}

To do so, we first apply the Baker––Campbell–Hausdorff expansion in Eq.~\eqref{eq:BCHLemma}, to the term $e^{\delta K_-}K_0 e^{-\delta K_-}$. Due to $[K_-, K_0] = K_-$, and $[K_-, [K_-, K_0]]=0$, we have
\begin{align}
    e^{\delta K_-}K_0 e^{-\delta K_-}&=\sum_{n=0}^\infty \frac{\delta^{n}[K_-, K_0]_{n}}{n!}= K_0 + \delta [K_-, K_0] = K_0 + \delta K_-.\label{eq:eK0e}
\end{align} 
In this case, Eq.\eqref{eq:Vk0V} becomes $e^{\beta K_+} e^{\gamma K_0}  (K_0 + \delta K_-) e^{-\gamma K_0} e^{-\beta K_+}$. Since $K_0$ commutes with itself, we have $e^{\gamma K_0} K_0 e^{-\gamma K_0} = K_0$. For $e^{\gamma K_0} \delta K_- e^{-\gamma K_0}$, due to $[K_0, K_-]=-K_-$ and $[K_0, [K_0, K_-]]= -[K_0, K_-]= (-1)^{2}K_-$, we can get 
\begin{align}
e^{\gamma K_0} \delta K_- e^{-\gamma K_0}&=\delta \sum_{n=0}^\infty \frac{\gamma^{n}[K_0, K_-]_{n}}{n!}=\delta \sum_{n=0}^\infty \frac{(-1\gamma)^n}{n!} K_- = \delta e^{-\gamma}K_-  \nonumber
\end{align} 

Therefore, we have 
\begin{align}
    e^{\gamma K_0} (K_0 + \delta K_-) e^{-\gamma K_0}= K_0 + \delta e^{-\gamma} K_- \label{eq:k0ek0}.
\end{align}

Now the Eq.\eqref{eq:Vk0V} becomes $e^{\beta K_+}(K_0 + \delta e^{-\gamma} K_-)e^{-\beta K_+}$. Since $[K_+, K_0]= -K_+$ and $[K_+, [K_+, K_0]]=0$, one can get
\begin{equation*}
    e^{\beta K_+}K_0 e^{-\beta K_+} = K_0 + \beta [K_+, K_0] = K_0 - \beta  K_+.
\end{equation*} 
Similarly, due to $[K_+, K_-]= -2K_0$, $[K_+, [K_+, K_-]]=-2 [K_+, K_0]= 2K_+$, and $[K_+, 2K_+]=0$, we have 
\begin{align}
e^{\beta K_+}(\delta e^{-\gamma}K_-)e^{-\beta K_+}=\delta e^{-\gamma} \sum_{n=0}^\infty \frac{\beta^{n}[K_+, K_-]_{n}}{n!}= e^{-\gamma}(K_- - 2\beta K_0 + \beta^{2} K_+)  
\end{align}

So, Eq.~\eqref{eq:Vk0V} becomes
\begin{align}
VK_0 V^{-1}&=e^{\beta K_+}(K_0 + \delta e^{-\gamma} K_-)e^{-\beta K_+} = K_0 - \beta  K_+ + \delta e^{-\gamma}(K_- - 2\beta K_0 + \beta^{2} K_+) \nonumber\\
&= (1-2\beta\delta e^{-\gamma})K_0 + \delta e^{-\gamma}K_- - \beta (1-e^{\gamma}\beta\delta)K_+   \label{eq:vk0}
\end{align}

Similarly, we calculate $VK_- V^{-1}$, which is 
\begin{align}
VK_- V^{-1}=(e^{\beta K_+} e^{\gamma K_0} e^{\delta K_-})K_- (e^{-\delta K_-} e^{-\gamma K_0} e^{-\beta K_+})\label{eq:VkmV}  
\end{align}
and obtain the final expressions for Eq.\eqref{eq:VkmV}
\begin{align}
VK_- V^{-1}=-2\beta e^{-\gamma}K_0 + e^{-\gamma}K_- + e^{-\gamma}\beta^{2}K_+\label{eq:vkm}
\end{align}

With the above, Eq.~\eqref{eq:S2normalorder} become $e^{P}=V$, leading to
\begin{align}
e^{P}K_0 e^{-P}=VK_0 V^{-1}, \,\, \,\ e^{P}K_- e^{-P}=VK_- V^{-1}\label{eq:k-p}
\end{align}

Using Eq.~\eqref{eq:k0}-\eqref{eq:km} and Eq.~\eqref{eq:vk0}-\eqref{eq:vkm}, we can rewrite Eq.~\eqref{eq:k-p} as
\begin{align}
&\cosh(2\lambda)K_0-\frac{\sinh(2\lambda)(e^{i\phi} K_+ + e^{-i\phi}K_-)}{2}= (1-2\beta\delta e^{-\gamma})K_0 + \delta e^{-\gamma}K_- - \beta (1-e^{\gamma}\beta\delta)K_+\label{eq:k0pnew} \\ 
&-e^{i\phi}\sinh(2\lambda)K_0+ \cosh^{2}(\lambda)K_- + e^{2i\phi}\sinh^{2}(\lambda)K_+ = -2\beta e^{-\gamma}K_0 + e^{-\gamma}K_- + e^{-\gamma}\beta^{2}K_+\label{eq:k-pnew}
\end{align}

Therefore, we have the following relationships:
\begin{align}
&\cosh(2\lambda)=(1-2\beta\delta e^{-\gamma}), \,\, \cosh^{2}(\lambda) = e^{-\gamma},\,\, \frac{e^{i\phi}\sinh(2\lambda)}{2} = \beta (1-e^{\gamma}\beta\delta)\nonumber\\
&e^{i\phi}\sinh(2\lambda)=2\beta e^{-\gamma},\,\,\, e^{2i\phi}\sinh^{2}(\lambda)= e^{-\gamma}\beta^{2} ,\,\,\,  -\frac{e^{-i\phi}\sinh(2\lambda) }{2}= \delta e^{-\gamma}\nonumber\\\nonumber
\end{align}
We now relate the coefficients \( \beta, \gamma, \delta \) to the parameters \( \lambda = r \) and \( \phi = \theta + \pi \) and one can have
\begin{align}
&e^{-\gamma} = \cosh^2(\lambda), \,\,\,\, e^{i\phi} \sinh(2\lambda) 
= 2\beta e^{-\gamma}, \,\,\,\, \delta e^{-\gamma} = -\frac{\sinh(2\lambda)e^{-i\phi}}{2}.\label{eq:delta_coef} 
\end{align}
This gives
\begin{align}
\gamma = -2 \ln (\cosh \lambda) = -2 \ln (\cosh r), \,\,\,\,
\beta = e^{i\phi} \tanh \lambda = -e^{i\theta} \tanh r,  \,\,\,\, \delta = -e^{-i\phi} \tanh \lambda = e^{-i\theta} \tanh r. 
\end{align}

In the end, we obtain Eq.~\eqref{eq:S2normal_order} in the main text
\begin{align}
S_2(\zeta)&=\exp \left(\zeta^* a b  - \zeta a^\dagger b^\dagger\right)=e^{\tau K_+ - \tau^* K_-}= e^{\beta K_+} e^{\gamma K_0} e^{\delta K_-} \\
&=\exp \left( -e^{i\theta} \tanh r \, a^\dagger b^\dagger \right)\times\exp \left( -\ln(\cosh r) \, (a^\dagger a + b^\dagger b + 1) \right)\times\exp \left(e^{-i\theta} \tanh r \, ab \right)\nonumber
\end{align}

\onecolumngrid
\subsection*{Analytical expression of squeezing state in Fock space}
Now we show the detailed steps to get the final quantum state when we apply $S_2 (\zeta)$ in the form of Eq.~\eqref{eq:S2normal_order} to the general Fock state $\ket{p,q}$, which is
\begin{align}
\ket{\psi}&=S_2(\zeta)\ket{p, q}\nonumber\\
&=\exp \left( -e^{i\theta} \tanh r \, a^\dagger b^\dagger \right)\exp \left( -\ln(\cosh r) \, (a^\dagger a + b^\dagger b + 1) \right)\exp \left(e^{-i\theta} \tanh r \, ab \right)\ket{p, q}\nonumber\\
&=\exp \left( -e^{i\theta} \tanh r \, a^\dagger b^\dagger \right)\exp \left( -\ln(\cosh r) \, (a^\dagger a + b^\dagger b + 1) \right)\sum_{n=0}^\infty \frac{(e^{-i\theta} \tanh r )^n}{n!} (ab)^n|p,q\rangle\nonumber\\
&=\exp \left( -e^{i\theta} \tanh r \, a^\dagger b^\dagger \right)\exp \left( -\ln(\cosh r) \, (a^\dagger a + b^\dagger b + 1) \right)\sum_{n=0}^{\min{(p,q)}} (e^{-i\theta} \tanh r)^n\,\sqrt{\binom{p}{n} \binom{q}{n}}\ket{p-n,q-n}.\nonumber\\
&=\exp \left( -e^{i\theta} \tanh r \, a^\dagger b^\dagger \right)\sum_{n=0}^{\min{(p,q)}} (e^{-i\theta} \tanh r)^n\,\sqrt{\binom{p}{n} \binom{q}{n}}\, \exp \left( -\ln(\cosh r) \, (a^\dagger a + b^\dagger b + 1) \right) \ket{p-n,q-n}\nonumber\\
&=\exp \left( -e^{i\theta} \tanh r \, a^\dagger b^\dagger \right)\sum_{n=0}^{\min{(p,q)}} (e^{-i\theta} \tanh r)^n\,\sqrt{\binom{p}{n} \binom{q}{n}}\times \sum_{k=0}^\infty \frac{(-\ln(\cosh r))^k}{k!}(a^\dagger a + b^\dagger b + 1)^k \ket{p-n,q-n}\nonumber\\
&=\exp \left( -e^{i\theta} \tanh r \, a^\dagger b^\dagger \right)\sum_{n=0}^{\min{(p,q)}} (e^{-i\theta} \tanh r)^n\,\sqrt{\binom{p}{n} \binom{q}{n}}\times \sum_{k=0}^\infty \frac{(-\ln(\cosh r))^k}{k!}(p+q-2n+1)^k \ket{p-n,q-n}\nonumber\\
&=\exp \left( -e^{i\theta} \tanh r \, a^\dagger b^\dagger \right)\sum_{n=0}^{\min{(p,q)}} (e^{-i\theta} \tanh r)^n\,\sqrt{\binom{p}{n} \binom{q}{n}} \exp \left( -\ln(\cosh r) \, (p+q-2n+1) \right) \ket{p-n,q-n}\nonumber\\
&=\exp \left( -e^{i\theta} \tanh r \, a^\dagger b^\dagger \right)\sum_{n=0}^{\min{(p,q)}} \frac{(e^{-i\theta} \tanh r)^n}{(\cosh r)^{p+q-2n+1}}\sqrt{\binom{p}{n} \binom{q}{n}} \ket{p-n,q-n}\nonumber\\
&=\sum_{n=0}^{\min{(p,q)}} \frac{(e^{-i\theta} \tanh r)^n}{(\cosh r)^{p+q-2n+1}}\sqrt{\binom{p}{n} \binom{q}{n}}\exp \left( -e^{i\theta} \tanh r \, a^\dagger b^\dagger \right) \ket{p-n,q-n}\nonumber\\
&=\sum_{n=0}^{\min{(p,q)}} \frac{(e^{-i\theta} \tanh r)^n}{(\cosh r)^{p+q-2n+1}}\sqrt{\binom{p}{n} \binom{q}{n}}\times \sum_{k=0}^\infty \frac{( -e^{i\theta} \tanh r )^k}{k!} (a^\dagger b^\dagger)^k\ket{p-n,q-n}\nonumber\\
&=\sum_{n=0}^{\min{(p,q)}} \frac{(e^{-i\theta} \tanh r)^n}{(\cosh r)^{p+q-2n+1}}\sqrt{\binom{p}{n} \binom{q}{n}}\times \sum_{k=0}^\infty ( -e^{i\theta} \tanh r )^k \sqrt{\binom{p-n+k}{k} \binom{q-n+k}{k}} \ket{p-n+k,q-n+k}\nonumber\\
&=\sum_{k=0}^\infty\sum_{n=0}^{\min{(p,q)}} \frac{(e^{-i\theta} \tanh r)^n ( -e^{i\theta} \tanh r )^k}{(\cosh r)^{p+q-2n+1}}\sqrt{\binom{p}{n} \binom{q}{n}\binom{p-n+k}{k} \binom{q-n+k}{k}} \ket{p-n+k,q-n+k} \label{detailedSPDC}
\end{align}

In the special case of  $p=q$, the Eq.\eqref{detailedSPDC} becomes
\begin{align}
S_2(\zeta)\ket{p,p}=\sum_{k=0}^\infty\sum_{n=0}^{p} \frac{(e^{-i\theta} \tanh r)^n ( -e^{i\theta} \tanh r )^k}{(\cosh r)^{2p-2n+1}}\binom{p}{n} \binom{p-n+k}{k} \ket{p-n+k,p-n+k}.
\end{align}
And for $p=q=0$ (applying $S_2(\zeta)$ to vacuum state), we have
\begin{align}
S_2(\zeta)\ket{0,0}=\sum_{k=0}^\infty \frac{( -e^{i\theta} \tanh r )^k}{\cosh r}\ket{k,k}
\end{align}

%Now, we define $m=q-n+k$, and $\Delta=p-q$, thus $\ket{p-n+k, q-n+k}$ becomes $\ket{m+\Delta, m}$, and $m=q-n+k\geq 0$ and $m+\Delta \geq 0$, $\min(m)$ is $\max(0,-\Delta)$. In the end, we have
%\begin{align}
%\ket{\psi}&=S_2(\zeta)\ket{p, q}=\sum_{\substack{m = \\ \max(0, -\Delta)}}^{\infty} C_{m,p,q}\,\, %\ket{m+\Delta, m}
%\end{align}
%with $C_{m,p,q}$ given by
%\begin{align}
%C_{m, p, q}&=\frac{(-e^{i\theta} \tanh r)^{m-q} \sqrt{p!q!(m+\Delta)!m!}}{(\cosh r )^{p+q+1}}\sum_{n=\max(0,q-m)}^{\min(p,q)} \frac{ (-1)^{n}(\sinh r)^{2n}}{n!(m-q+n)!(p-n)!(q-n)!}\nonumber
%\end{align}

\subsection*{Quantum state in the two-crystal setup}
The analytic form of the output quantum state for the two-crystal configuration in Fig.~\ref{2crystal} is given below. For clarity, we assume all nonlinear source phases are zero.
\begin{align}
\ket{\Psi}&=S_{ab}(r_{2})\Phi_a (\phi)S_{ab}(r_{1})\ket{0,0}\nonumber\\
&=S_{ab}(r_{2})\Phi_a (\phi) \sum_{k_{1}=0}^\infty \frac{(-\tanh r_{1} )^{k_{1}}}{\cosh r_{1}}\ket{k_{1},k_{1}}\nonumber\\
&=S_{ab}(r_{2})\sum_{k_{1}=0}^\infty \frac{(-\tanh r_{1} )^{k_{1}}\, e^{i\phi k_{1}}}{\cosh r_{1}}\ket{k_{1},k_{1}}\nonumber\\
&=S_{ab}(r_{2})\sum_{k_{1}=0}^\infty \frac{(-\tanh r_{1} )^{k_{1}}\, e^{i\phi k_{1}}}{\cosh r_{1}}\ket{k_{1},k_{1}}\nonumber\\
&=\sum_{k_{1}=0}^\infty \sum_{k_{2}=0}^\infty\sum_{n=0}^{k_{1}} \frac{e^{i\phi k_{1}}(-\tanh r_{1} )^{k_{1}}(\tanh r_{2})^n (-\tanh r_{2} )^{k_{2}}}{\cosh r_{1}\times (\cosh r_{2})^{2k_{1}-2n+1}}\binom{k_{1}}{n} \binom{k_{1}-n+k_{2}}{k_{2}} \ket{k_{1}-n+k_{2},\, k_{1}-n+k_{2}}
\end{align}

\subsection*{Quantum state in the three-crystal setup}
The analytic form of the output quantum state for the three-crystal configuration in Fig.~\ref{3crystal} is given below. For clarity, we assume all nonlinear source phases are zero.
\begin{align}
\ket{\Psi}&=S_{ab}(r_{3})\Phi_a (\phi_{2})S_{ab}(r_{2})\Phi_a (\phi_{1})S_{ab}(r_{1})\ket{0,0}\nonumber\\
&=S_{ab}(r_{3})\Phi_a (\phi_{2})S_{ab}(r_{2})\Phi_a (\phi_{1}) \sum_{k_{1}=0}^\infty \frac{(-\tanh r_{1} )^{k_{1}}}{\cosh r_{1}}\ket{k_{1},k_{1}}\nonumber\\
&=S_{ab}(r_{3})\Phi_a (\phi_{2})S_{ab}(r_{2})\sum_{k_{1}=0}^\infty \frac{(-\tanh r_{1} )^{k_{1}}\, e^{i\phi_{1} k_{1}}}{\cosh r_{1}}\ket{k_{1},k_{1}}\nonumber\\
&=S_{ab}(r_{3})\Phi_a (\phi_{2})\sum_{k_{1}=0}^\infty \sum_{k_{2}=0}^\infty\sum_{n=0}^{k_{1}} \frac{e^{i\phi_{1} k_{1}}(-\tanh r_{1} )^{k_{1}}(\tanh r_{2})^n (-\tanh r_{2} )^{k_{2}}}{\cosh r_{1}\times (\cosh r_{2})^{2k_{1}-2n+1}}\binom{k_{1}}{n} \binom{k_{1}-n+k_{2}}{k_{2}} \ket{k_{1}-n+k_{2},\, k_{1}-n+k_{2}}\nonumber\\
&=\sum_{k_{1}=0}^\infty \sum_{k_{2}=0}^\infty\sum_{n=0}^{k_{1}} \Bigg[ \frac{e^{i\phi_{1} k_{1}+i\phi_{2}( k_{1}-n+k_{2})}(-\tanh r_{1} )^{k_{1}}(\tanh r_{2})^n (-\tanh r_{2} )^{k_{2}}}{\cosh r_{1}\times (\cosh r_{2})^{2k_{1}-2n+1}}\binom{k_{1}}{n} \binom{k_{1}-n+k_{2}}{k_{2}}\nonumber\\
&\quad\times  S_{ab}(r_{3})\ket{k_{1}-n+k_{2},\, k_{1}-n+k_{2}}\Bigg]\nonumber\\
&=\sum_{k_{1}=0}^\infty \sum_{k_{2}=0}^\infty \sum_{k_{3}=0}^\infty \sum_{n=0}^{k_{1}} \sum_{m=0}^{k_{1}-n+k_{2}} \Bigg[ \frac{e^{i\phi_{1} k_{1}+i\phi_{2}( k_{1}-n+k_{2})}(-\tanh r_{1} )^{k_{1}}(\tanh r_{2})^n (-\tanh r_{2} )^{k_{2}}(\tanh r_{3})^m (-\tanh r_{3} )^{k_{3}}}{\cosh r_{1}\times (\cosh r_{2})^{2k_{1}-2n+1}\times (\cosh r_{3})^{2(k_{1}-n+k_{2})-2m+1}}\nonumber\\
&\quad\times \binom{k_{1}}{n}\binom{k_{1}-n+k_{2}}{k_{2}}\binom{k_{1}-n+k_{2}}{m}\binom{k_{1}-n+k_{2}-m+k_{3}}{k_{3}}\ket{k_{1}-n+k_{2}-m+k_{3},\, k_{1}-n+k_{2}-m+k_{3}}\Bigg]
\end{align}

\subsection*{Quantum state in the four-crystal setup}
The analytic form of the output quantum state for the three-crystal configuration in Fig.~\ref{4crystal} is given below. For clarity, we assume all nonlinear source phases are zero.
\begin{align}
\ket{\Psi}&=S_{cd}(r_{4})S_{ab}(r_{3})\Phi_a (\phi)S_{bd}(r_{2})S_{ac}(r_{1})\ket{0,0,0,0}_{abcd}\nonumber\\
&=S_{cd}(r_{4})S_{ab}(r_{3})\Phi_a (\phi)S_{bd}(r_{2})\sum_{k_{1}=0}^\infty \frac{(-\tanh r_{1} )^{k_{1}}}{\cosh r_{1}}\ket{k_{1},0, k_{1}, 0}\nonumber\\
&=S_{cd}(r_{4})S_{ab}(r_{3})\Phi_a (\phi)\sum_{k_{1}=0}^\infty \sum_{k_{2}=0}^\infty \frac{(-\tanh r_{1} )^{k_{1}}(-\tanh r_{2} )^{k_{2}}}{\cosh r_{1} \cosh r_{2}}\ket{k_{1}, k_{2}, k_{1}, k_{2}}\nonumber\\
&=S_{cd}(r_{4})S_{ab}(r_{3})\sum_{k_{1}=0}^\infty \sum_{k_{2}=0}^\infty \frac{e^{i\phi k_{1}}(-\tanh r_{1} )^{k_{1}}(-\tanh r_{2} )^{k_{2}}}{\cosh r_{1} \cosh r_{2}}\ket{k_{1}, k_{2}, k_{1}, k_{2}}\nonumber\\
&=S_{cd}(r_{4})\sum_{k_{1}=0}^\infty \sum_{k_{2}=0}^\infty \sum_{k_{3}=0}^\infty\sum_{n=0}^{\min{(k_{1}, k_{2})}} \Biggl[ \frac{e^{i\phi k_{1}}(-\tanh r_{1} )^{k_{1}}(-\tanh r_{2} )^{k_{2}}}{\cosh r_{1} \cosh r_{2}}\times \frac{(\tanh r_{3})^n ( -\tanh r_{3} )^{k_{3}}}{(\cosh r_{3})^{k_{1}+k_{2}-2n+1}}\nonumber\\
&\quad\times \sqrt{\binom{k_{1}}{n} \binom{k_{2}}{n}\binom{k_{1}-n+k_{3}}{k_{3}} \binom{k_{2}-n+k_{3}}{k_{3}}} \ket{k_{1}-n+k_{3},\,\, k_{2}-n+k_{3},\,\, k_{1},\,\, k_{2}}\Biggl]\nonumber\\
&=\sum_{k_{1}=0}^\infty \sum_{k_{2}=0}^\infty \sum_{k_{3}=0}^\infty\sum_{k_{4}=0}^\infty \sum_{n=0}^{\min{(k_{1}, k_{2})}} \sum_{m=0}^{\min{(k_{1}, k_{2})}}\nonumber \\
& \quad \Biggl[\frac{e^{i\phi k_{1}}(-\tanh r_{1} )^{k_{1}}(-\tanh r_{2} )^{k_{2}}}{\cosh r_{1} \cosh r_{2}}\times \frac{(\tanh r_{3})^n ( -\tanh r_{3} )^{k_{3}}}{(\cosh r_{3})^{k_{1}+k_{2}-2n+1}}\times \frac{(\tanh r_{4})^m ( -\tanh r_{4} )^{k_{4}}}{(\cosh r_{4})^{k_{1}+k_{2}-2m+1}}\nonumber\\
&\quad\times \sqrt{\binom{k_{1}}{n} \binom{k_{2}}{n}\binom{k_{1}-n+k_{3}}{k_{3}} \binom{k_{2}-n+k_{3}}{k_{3}}\binom{k_{1}}{m} \binom{k_{2}}{m}\binom{k_{1}-m+k_{4}}{k_{4}} \binom{k_{2}-m+k_{4}}{k_{4}}}\nonumber\\
&\quad\times \ket{k_{1}-n+k_{3},\,\, k_{2}-n+k_{3},\,\, k_{1}-m+k_{4},\,\, k_{2}-m+k_{4}}\Biggl]\nonumber\\
\end{align}

\end{document}